\documentclass[superscriptaddress, twocolumn, prx, longbibliography, nofootinbib]{revtex4-1}
\usepackage{amsmath, amssymb, color, graphicx, stmaryrd, wasysym, esint}
\usepackage[utf8]{inputenc}
\definecolor{linkcolor}{rgb}{0,0,0.6} 
\usepackage[pdftex,colorlinks=true,
	pdfstartview = FitV,
	linkcolor    = linkcolor,
	citecolor    = linkcolor,
	urlcolor     = linkcolor,	
	hyperindex   = true,
	hyperfigures = false]{hyperref}

\begin{document}

\title{Thermodynamics of active field theories: Energetic cost of coupling to reservoirs}

\author{Tomer Markovich}
\affiliation{Center for Theoretical Biological Physics, Rice University, Houston, Texas 77030, USA}
\affiliation{DAMTP, Centre for Mathematical Sciences, University of Cambridge, Wilberforce Road, Cambridge CB3 0WA, UK}

\author{\'Etienne Fodor}
\affiliation{DAMTP, Centre for Mathematical Sciences, University of Cambridge, Wilberforce Road, Cambridge CB3 0WA, UK}
\affiliation{Department of Physics and Materials Science, University of Luxembourg, L-1511 Luxembourg}

\author{Elsen Tjhung}
\affiliation{DAMTP, Centre for Mathematical Sciences, University of Cambridge, Wilberforce Road, Cambridge CB3 0WA, UK}
\affiliation{Department of Physics, University of Durham, Science Laboratories, South Road, Durham DH1 3LE, UK}

\author{Michael E. Cates}
\affiliation{DAMTP, Centre for Mathematical Sciences, University of Cambridge, Wilberforce Road, Cambridge CB3 0WA, UK}

\begin{abstract}
	The hallmark of active matter is the autonomous directed motion of its microscopic constituents driven by consumption of energy resources. This leads to the emergence of large scale dynamics and structures without any equilibrium equivalent. Though active field theories offer a useful hydrodynamic description, it is unclear how to properly quantify the energetic cost of the dynamics from such a coarse-grained description. We provide a thermodynamically consistent framework to identify the energy exchanges between active systems and their surrounding thermostat at the hydrodynamic level. Based on linear irreversible thermodynamics, we determine how active fields couple with the underlying reservoirs at the basis of nonequilibrium driving. This leads to evaluating the rate of heat dissipated in the thermostat, as a measure of the cost to sustain the system away from equilibrium, which is related to the irreversibility of the active field dynamics. We demonstrate the applicability of our approach in two popular active field theories: (i)~the dynamics of a conserved density field reproducing active phase separation, and (ii)~the coupled dynamics of density and polarization describing motile deformable droplets. Combining numerical and analytical approaches, we provide spatial maps of dissipated heat, compare them with the irreversibility measure of the active field dynamics, and explore how the overall dissipated heat varies with the emerging order.
\end{abstract}

\maketitle


\section{Introduction}

Active materials are those in which each component extracts energy from the environment to produce a directed motion~\cite{Marchetti2013, Bechinger2016, Marchetti2018}. Examples of active systems can be found in the realm of living matter, such as swarms of bacteria~\cite{Kessler2004, Goldstein2007, Aranson2012}, cells~\cite{Ladoux2017, Sano2017, fred2010, Chen2020} and bird flocks~\cite{Bialek2012, Cavagna2014}, but also in reconstituted or biomimetic realizations, such as motility assays~\cite{Sanchez2012, Decamp2015}, Janus particles in a fuel bath~\cite{Golestanian2007, Bechinger2013, Palacci2013} and vibrated polar particles~\cite{Narayan2007, Dauchot2010}. To explore the collective effects emerging at large scales, several studies have focused on minimal models which reproduce, for instance, the macroscopic collective motion between aligning particles~\cite{Vicsek1995, Toner1995} and the clustering between purely repulsive agents~\cite{Tailleur2008, Cates2015}. Such theories can be particle-based, thus relying on postulating the form of nonequilibrium forces at the micro-scale, or given by hydrodynamic equations involving fluctuating fields. In the latter case, the dynamics are either obtained from a systematic coarse-graining procedure~\cite{Bertin2009, Farrell2012, Speck2013, Tailleur2013, Markovich2020} or directly postulated based on phenomenological arguments~\cite{Toner1995, Wittkowski2014, Nardini2017, Nardini2018, Rapp2019, Markovich2019, Markovich2019b}.

In recent years, a large number of works have focused on developing a thermodynamic approach to active matter. They are led by the search for generic observables to quantify, classify and predict the anomalous properties of active systems. For instance, the pressure and chemical potential allow one to distinguish systems depending on whether or not they obey equations of state~\cite{Takatori2015, Solon2015b, Paliwal2018, Guioth2019}, and each can be useful to predict phase diagrams~\cite{Solon2018, Solon2018b}. Moreover, quantifying the irreversibility of the dynamics enables us to locate where and when activity mainly affects the system~\cite{Nardini2016, Nardini2017, Murrell2019}, and to explore the relation between irreversibility and phase transitions~\cite{Shim2016, Spinney2018}. This has motivated several experiments to measure the dissipation associated with irreversibility in various systems~\cite{Battle2016, Murrell2018, Mura2018, Roichman2018}. Furthermore, it has been shown that, for minimal active models, changing the dissipation by using a dynamical bias provides an alternative route to clustering and collective motion in active matter~\cite{Suma2017, Nemoto2018, Suri2019, Suri2020, GrandPre2020}.

Progress in building the thermodynamics of active matter has been mainly achieved so far in particle-based descriptions. Indeed, such dynamics bear a natural mechanical interpretation of energy exchanges, with either an external operator or the surrounding thermostat, in terms of forces and displacements. The tools of stochastic thermodynamics, first introduced in thermal systems~\cite{Sekimoto1998, Seifert2012} and then extended to active ones~\cite{Speck2016, Nardini2016, Mandal2017, Maggi2017, Shankar2018, Bo2019, Bisker2019}, offer a framework to quantify systematically work, heat, and entropy production from the microscopic dynamics. They also allow one to describe the consumption of chemical fuel, at the basis of self-propulsion, in a thermodynamically consistent manner~\cite{Seifert2018, Speck2018, Kapral2018}. In contrast, though some hydrodynamic approaches consider coupling with a momentum-conserving fluid~\cite{Marchetti2013, Yeomans2013, Giomi2015, Hemingway2015}, nonequilibrium terms in many field dynamics do not rely on any explicit mechanical force~\cite{Toner1995, Wittkowski2014, Nardini2017, Nardini2018, Rapp2019}. Then, a systematic definition of how active systems exchange energy with their environment at a coarse-grained level has been elusive: It remains to build the energetics of active field theories.

A major breakthrough of stochastic thermodynamics is to relate explicitly the irreversibility of the dynamics, as measured by the divergence of forward and time-reversed realizations, with the amount of energy dissipated in the thermostat~\cite{Sekimoto1998, Seifert2012}. This connection only holds for thermodynamically consistent dynamics, whose formulation is constrained so that the connection to the underlying thermostat is properly taken into account. Importantly, the active field theories postulated only from symmetry arguments do not generally satisfy these constraints a priori~\cite{Toner1995, Wittkowski2014, Nardini2017, Nardini2018, Rapp2019}. Hence, it is unclear to which extent the measure of irreversibility in these models, often referred to as entropy production rate (EPR) and already evaluated in previous works~\cite{Nardini2017, Ramaswamy2018, Murrell2019}, actually provides relevant information about energy dissipation.

Interestingly, linear irreversible thermodynamics (LIT) provides a definition of dissipation in terms of the thermodynamic forces and conjugated currents at a hydrodynamic level~\cite{Mazur}. After identifying the relevant forces and currents for a given theory, the field dynamics are formulated by postulating linear relations between them. These theories were originally designed to capture the effect of external drives, such as temperature gradients or electric fields, yet extensions to systems with internal activity, such as {\it active gels}, have been shown to successfully reproduce the behavior of living materials~\cite{Kruse2004, Joanny2009, Prost2017}. Though some active theories do not follow linear force-current relations a priori, it is tempting to draw analogies with LIT in order to systematically define dissipation in this broad class of dynamics beyond active gels. The challenge is then to embed active field theories with arbitrary nonequilibrium terms into the specific structure of LIT, thus enforcing a thermodynamically consistent framework.
This is a non-trivial task with the benefit of drawing powerful generic results from thermodynamic considerations.

In what follows, we offer a framework to evaluate the energetic budget of active field theories. Starting from first principles, we demonstrate how to define the heat rate dissipated to the thermostat from the fluctuations of the active fields. Importantly, we show that the heat rate can be generically decomposed into a homogeneous background contribution, independent of active fields, and a contribution given in terms of the statistics of the active fields. This decomposition allows one to quantify how the structure and dynamics of active fields affect where heat is dissipated, thus opening the door to estimating and comparing the energetic cost associated with different emerging orders. To illustrate the relevance of our framework, we apply it to field theories which capture the emergence of a phase separation and/or a polar order. Overall, our results demonstrate the ability to estimate the rate of energy required to sustain a given active dynamics away from equilibrium.

Our approach relies on systematically constructing the dynamics of a set of underlying fields, which drive the system out of equilibrium, from that of the active field dynamics based on the force-current relations of LIT. Importantly, under non-restrictive assumptions (see Sec.~\ref{sec:dyn}), the evolution of active fields remains independent of that of the driving fields: The latter are hidden degrees of freedom that do not affect the emerging order. This leads to show that the heat rate, whose expression follows from the total EPR measuring the irreversibility of both the active and driving fields, can be evaluated from the fluctuations of active fields only. Importantly, the heat rate is distinct from the EPR quantifying the irreversibility of the active field dynamics alone, which we refer to as the {\it explicit} EPR in what follows.

We analyze in detail the heat rate in two popular models for active matter: (i)~the dynamics of active phase separation, known as Active Model B~\cite{Wittkowski2014, Nardini2017}, and (ii)~the dynamics of polar motile droplets~\cite{Tjhung2012, Tjhung2015}. For model (i), we find that the heat-rate mainly varies at the interfaces between dense and dilute phases, where it reduces compared to its bulk value. We further evaluate the heat-rate scaling with noise amplitude and driving parameter. Our results are compared with the explicit EPR, as an alternative measure of the deviation from equilibrium~\cite{Nardini2017, Ramaswamy2018, Murrell2019}. The analysis of model (ii) reveals that the rate  of dissipated heat varies across the profile of polar droplets. We also report a hysteresis loop associated with the splitting and fusion of multiple droplets, and we discuss the scaling with noise amplitude and driving parameter.

The paper is organized as follows. First, we present in Sec.~\ref{sec:frame} how to embed generic active field theories within LIT and calculate the heat rate, which we then relate with the explicit EPR in Sec.~\ref{sec:app}. In Sec.~\ref{sec:phase}, we consider an application of our framework to dynamics that capture active phase separation in terms of a conserved scalar field~\cite{Wittkowski2014, Nardini2017}. To go beyond scalar field theories, we then study in Sec.~\ref{sec:drop} the dynamics of motile droplets coupling density and polarization~\cite{Tjhung2012, Tjhung2015} as a prototypical model of deformable living cells~\cite{Camley2014, Palmieri2015, Yeomans2019}. Finally, we present our conclusions in Sec.~\ref{sec:conclusion}.


\section{The role of underlying reservoirs}\label{sec:frame}

\begin{figure}
	\centering
	\includegraphics[width=\linewidth]{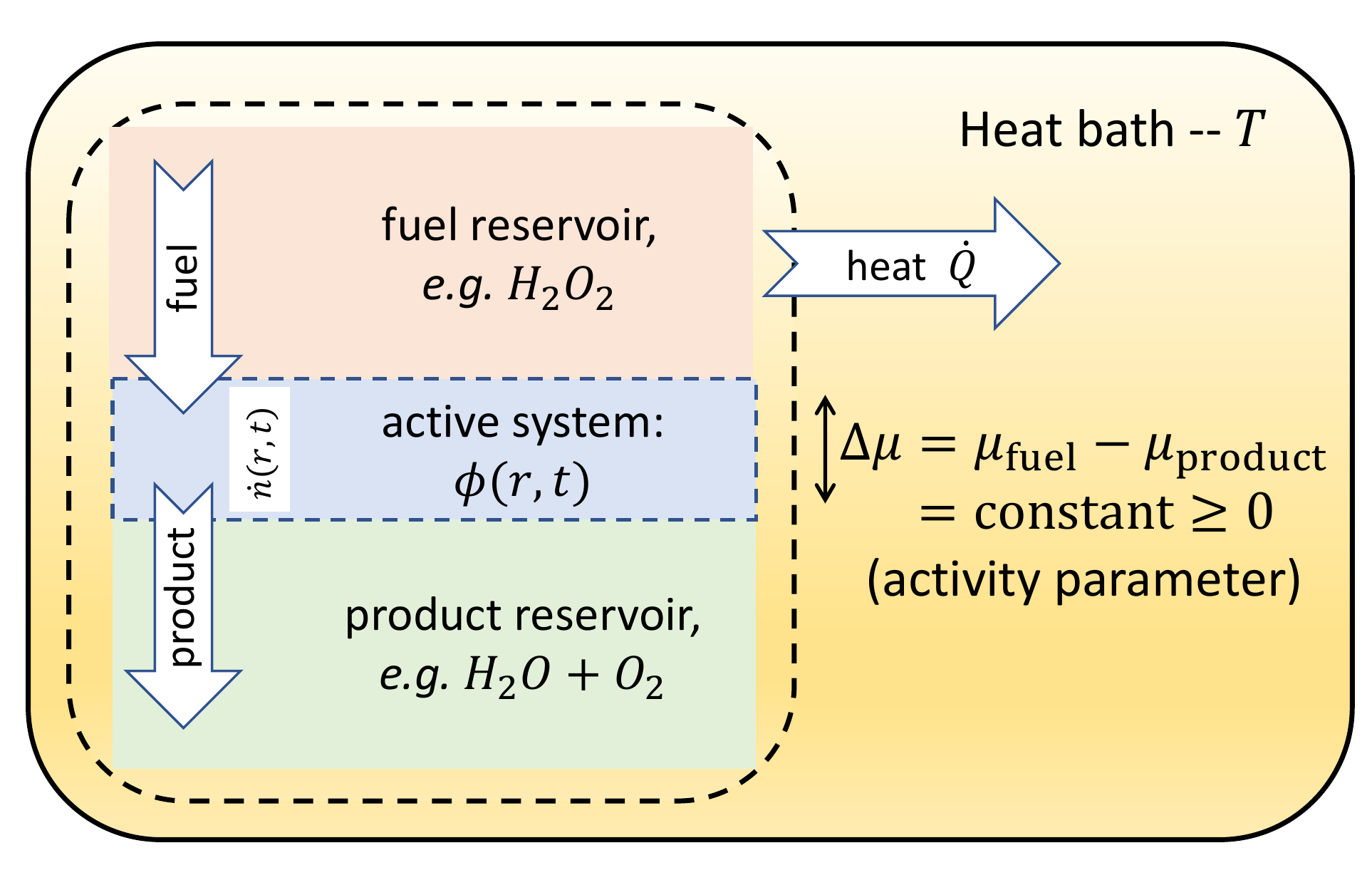}
	\caption{\label{fig1}
		Schematic representation of an active system (blue) put in contact with reservoirs of chemical fuel (red) and product (green) which set a constant, homogeneous chemical potential difference $\Delta\mu$ in the active system. This is essentially a nonequilibrium grand-canonical ensemble for the active system (details in Appendix~\ref{app:grand}).
		Within our framework, $\Delta\mu$ embodies the driving parameter which controls the nonequilibrium terms in the dynamics (\ref{eq:dyn_phi}-\ref{eq:dyn_chem}) for the active density field $\phi$ and the rate of fuel consumption $\dot{n}$. The active system and the chemical reservoirs are surrounded by the thermostat (yellow) which maintains a fixed temperature $T$. The fluctuations of $\phi$ and $n$ lead to dissipation of heat ${\cal Q}$ into the thermostat, which quantifies the energetic cost to maintain the whole system away from equilibrium.
	}
\end{figure}

We consider active dynamics of hydrodynamic fields which can be either obtained from explicit coarse-graining of microscopic dynamics, or written phenomenologically using symmetry arguments~\cite{Marchetti2013, Marchetti2018}. Our approach consists in introducing additional fields, associated for instance with chemical reactions which sustain the dynamics away from equilibrium, see Fig.~\ref{fig1}. This amounts to identifying the nonequilibrium terms in the original dynamics as a coupling to chemical reservoirs following the framework of linear irreversible thermodynamics~\cite{Mazur}. 
Below, we present in detail the procedure to enforce a  thermodynamically consistent structure of the dynamics: First, for a conserved scalar field, and then for generalized field dynamics which couple a conserved scalar field and a polarization field.
The key in providing a thermodynamic framework for active materials is to realize that such materials are typically a part of a larger system, which provides the drive needed to sustain nonequilibrium activity, as described in Fig.~\ref{fig1}.

\subsection{Coupling active and chemical fields}\label{sec:dyn}

To introduce pedagogically our framework, we start by considering the dynamics of a conserved scalar field $\phi$ representing the density of active components:
\begin{equation}\label{eq:dyn_phi}
	\dot\phi = - \nabla\cdot{\bf J} ,
	\quad
	{\bf J} = - \lambda\nabla\frac{\delta\cal F}{\delta\phi} + \Delta\mu\,{\bf C} + T \,{\boldsymbol\nu}({\bf C}) + {\boldsymbol\Lambda} ,
\end{equation}
where $\cal F$ is the free energy, $\lambda$ is the mobility, $\Delta\mu$ is the driving coefficient, and ${\bf C}$ is a vector-valued function of $\phi$ and its gradients. The noise term $\boldsymbol\Lambda$ is Gaussian with zero mean and correlations given by
\begin{equation}\label{eq:lambda}
	\big\langle\Lambda_\alpha({\bf r},t)\Lambda_\beta({\bf r}',t')\big\rangle = 2\lambda T \delta_{\alpha\beta}\delta({\bf r}-{\bf r}')\delta(t-t') , 
\end{equation}
where $T$ is the temperature of the surrounding heat bath. The term $T{\boldsymbol\nu}$ is a generalization of the spurious drift that typically appears in ordinary stochastic differential equations with multiplicative noise. Its expression is determined by that of ${\bf C}$, it depends on both time and space discretizations, and it obviously vanishes when fluctuations are neglected ($T=0$). 
In Appendix~\ref{app:discretisation} we generalize standard results for stochastic processes~\cite{Gardiner} to stochastic field theories and derive the expression for the spurious drift term.
The dynamics~\eqref{eq:dyn_phi} has been used extensively to reproduce the phase separation of active particles~\cite{Speck2013, Tailleur2013, Wittkowski2014, Nardini2017, Nardini2018, Rapp2019}. In these works, the need for the additional term $T\boldsymbol\nu$ was not addressed explicitly, mainly because previous studied were not concerned with thermodynamic consistency, and also since the noise $\boldsymbol\Lambda$ seems to be additive when considering only the fluctuations of $\phi$. When we describe below the origin of $\Delta\mu$, by introducing additional field dynamics, it will become apparent that the noise $\boldsymbol\Lambda$ is in fact multiplicative due to its cross-correlation with the noise of the additional field, as described in Eq.~\eqref{eq:dyn_chem}.

Our goal is to connect the emergent behavior of $\phi$ with the underlying consumption of energy resources. To this end, we describe explicitly the fluctuations of the degrees of freedom at the basis of nonequilibrium drive, referred to as {\it chemical fields} in what follows, though our framework extends to other types of drive. Inspired by recent works~\cite{Ramaswamy2017, Ramaswamy2018}, we regard the driving coefficient $\Delta\mu$ as the chemical potential difference between fuel and products of a chemical reaction, see Fig.~\ref{fig1}, which applies, for instance, to the oxidation of hydrogen peroxide involved in the self-propulsion of Janus colloids~\cite{Golestanian2007, Bechinger2013, Palacci2013}. This leads us to consider the dynamics of the chemical coordinate $n$, which is (half) the difference between the local number density of product molecules and that of the fuel molecules (see Appendix~\ref{app:grand}). It is described as a field fluctuating in space and time, while $\Delta\mu$ is kept constant and homogeneous.

We aim at proposing a systematic method to couple the active field $\phi$ and the underlying chemical field $n$. It relies on the fact that the active system is a part of a large nonequilibrium system that relaxes (slowly) towards equilibrium. With this assumption, the explicit dynamics of $n$ can be deduced from linear irreversible thermodynamics (LIT)~\cite{Mazur, Basu2008, Prost2017, Julicher2018, Markovich2019b}. Identifying ${\bf J}$ and $-\nabla(\delta{\cal F}/\delta\phi)$ as the current and the thermodynamic force associated with $\phi$, respectively, LIT states that the currents $\{{\bf J}, \dot n\}$ can be written as a linear combination of the thermodynamic forces $\{-\nabla(\delta{\cal F}/\delta\phi), \Delta\mu\}$. It is clear from~\eqref{eq:dyn_phi} that the factor coupling the current $\bf J$ and the force $\Delta\mu$ is directly given by ${\bf C}$. Accordingly, and because $\phi$ is even under time-reversal, Onsager reciprocity relations require that the coupling factor between the current $\dot n$ and the force $-\nabla(\delta{\cal F}/\delta\phi)$ is also $\bf C$~\cite{Onsager1931}, so that the dynamics of $n$ follows as
\begin{equation}\label{eq:dyn_chem_g}
	\dot n = \gamma\Delta\mu - {\bf C}\cdot\nabla\frac{\delta\cal F}{\delta\phi} + T \,\chi({\bf C}) + \xi ,
\end{equation}
where $\gamma$ is the chemical mobility, which we take constant in what follows. As a result of this assumption, the equation for $\phi$ is autonomous and does not rely on knowing the fluctuations of the chemical field $n$. The noise term $\xi$ is Gaussian with zero mean and correlations given by
\begin{equation}\label{eq:dyn_chem}
	\begin{aligned}
		\big\langle\xi({\bf r},t)\xi({\bf r}',t')\big\rangle &= 2\gamma T\delta({\bf r}-{\bf r}')\delta(t-t') ,
		\\
		\big\langle\Lambda_\alpha({\bf r},t)\xi({\bf r}',t')\big\rangle &= 2 T C_\alpha({\bf r},t)\delta({\bf r}-{\bf r}')\delta(t-t') .
	\end{aligned}
\end{equation}
Note that, though LIT states linear relations between forces and currents, the coupling factor $\bf C$ need not be linear with respect to $\phi$ and its gradients.

It is convenient to introduce the Onsager matrix $\mathbb L$ which gives the coupling between forces and currents in $d+1$ dimensions. For $d=3$, it is given by
\begin{equation}\label{eq:L}
	{\mathbb L} =
	\begin{bmatrix}
		\lambda & 0 & 0 & C_x
		\\
		0 & \lambda & 0 & C_y
		\\
		0 & 0 & \lambda & C_z
		\\
		C_x & C_y & C_z & \gamma
	\end{bmatrix}
	.
\end{equation}
Then, the dynamics~(\ref{eq:dyn_phi}-\ref{eq:dyn_chem}) can be expressed in a compact form as~\cite{Chaikin}
\begin{equation}\label{eq:onsager}
 \big[{\bf J},\dot n\big] = {\mathbb L} \bigg[ -\nabla\frac{\delta\cal F}{\delta\phi}, \Delta\mu\bigg] + T \,\big[{\boldsymbol\nu},\chi\big] + \big[{\boldsymbol\Lambda}, \xi\big] ,
\end{equation}
where the noise correlations read 
\begin{equation}
	\big\langle \big[{\boldsymbol\Lambda}, \xi\big]({\bf r},t) \big[{\boldsymbol\Lambda}, \xi\big]^\intercal({\bf r}',t') \big\rangle = 2T\,{\mathbb L}({\bf r},t)\delta({\bf r}-{\bf r}')\delta(t-t'),
\end{equation}
and $^\intercal$ denotes transpose. The expression of $\{{\boldsymbol\nu},\chi\}$ can be obtained from that of the Onsager matrix $\mathbb L$ following a systematic route, as detailed in Appendix~\ref{app:discretisation}. In particular, it depends on the choice of how the gradient terms appearing in $\bf C$ are discretized in space, see Appendix~\ref{app:discretisation}. In the specific examples considered below, a judicious choice of the discretization can be made such that the spurious drift vanishes. Moreover, one can show that ${\boldsymbol\nu}={\bf 0}$ for $d=1$, and that $\{{\boldsymbol\nu},\chi\}$ both vanish whenever $\bf C$ is a local function of $\phi$ independent of its gradients.

The dynamics~(\ref{eq:dyn_phi}-\ref{eq:dyn_chem}) is thermodynamically consistent in the sense that it obeys detailed balance, and thus relaxes to an equilibrium state at temperature $T$, when $\Delta\mu$ derives from a given chemical free energy ${\cal F}_{\rm ch}$ so that $\Delta\mu = -\delta{\cal F}_{\rm ch}/\delta n$~\cite{Mazur, Basu2008, Prost2017, Julicher2018}, see Appendix~\ref{app:grand}. Equilibrium relaxation also requires that the Onsager matrix is positive semi-definite ($\det{\mathbb L}\geq0$). 
When considering the dynamics within the active system, $\Delta\mu$ can be regarded as constant, see the nonequilibrium grand-canonical ensemble described in Appendix~\ref{app:grand}. Then, the realizations of the active field $\phi$ are independent of that of the chemical field $n$, and the dynamics now operates away from equilibrium. Although the realizations of $\phi$ are independent of $n$, the presence of $n$ determines the existence of the spurious drift term $\boldsymbol{\nu}$ and thus affects the $\phi$ dynamics. Within this grand-canonical description, $\dot{n}$ is the important field (rather than $n$) and it should be thought of as the local rate of chemical reactions.


\subsection{Dissipation and irreversibility}\label{sec:entropy}

The nonequilibrium drive $\Delta\mu$ breaks time-reversal symmetry and leads to dissipation of energy in the form of heat $\cal Q$ from the system to the surrounding thermostat. Following stochastic thermodynamics~\cite{Lebowitz1999, Sekimoto1998, Seifert2012}, the heat along a trajectory can be evaluated from the irreversibility of the dynamics.  It amounts to comparing the path probabilities of the forward and time-reversed dynamics, respectively denoted $\cal P$ and ${\cal P}^{\rm R}$, which quantify the probability of observing a trajectory of the currents $\{{\bf J}, \dot n\}$ within a given time interval $[0,t]$\footnote{In the presence of multiplicative noise, writing explicitly the path probabilities in~\eqref{eq:traj_heat} and \eqref{eq:entropy} requires a careful treatment~\cite{Lau2007, Spinney2012, Lecomte2017}. We use here mid-point temporal discretization with Stratonovich convention for the forward and backward trajectories.}:
\begin{equation}\label{eq:traj_heat}
{\cal Q} = T \ln \frac{{\cal P}\big[\{{\bf J}, \dot n\}_0^t\big]}{{\cal P}^{\rm R}\big[\{{\bf J}, \dot n\}_0^t\big]} \, .
\end{equation}
The steady state heat rate $\dot{\cal Q}$ is then
\begin{equation}\label{eq:entropy}
	\dot{\cal Q} = T \left< \underset{t\to\infty}{\lim} \frac{1}{t} \ln \frac{{\cal P}\big[\{{\bf J}, \dot n\}_0^t\big]}{{\cal P}^{\rm R}\big[\{{\bf J}, \dot n\}_0^t\big]} \right> ,
\end{equation}
where the average is taken with respect to noise realization (or ${\cal P}\big[\{{\bf J}, \dot n\}_0^t\big]$).
In equilibrium, the dynamics are symmetric under time reversal with the same statistics for forward and backward trajectories (${\cal P}={\cal P}^{\rm R}$), so that the system does not dissipate any heat ($\dot{\cal Q}=0$). In the presence of nonequilibrium drive in steady state, time-reversal symmetry is broken (${\cal P}\neq{\cal P}^{\rm R}$) which yields a constant rate of dissipation in steady state ($\dot{\cal Q}>0$).

The irreversibility of the dynamics can also be evaluated at the level of active field alone:
\begin{equation}
	{\cal S} = \left< \underset{t\to\infty}{\lim} \frac{1}{t} \ln \frac{{\cal P}\big[\{{\bf J}\}_0^t\big]}{{\cal P}^{\rm R}\big[\{{\bf J}\}_0^t\big]} \right>.
\end{equation}
The irreversibility measure $\cal S$, referred to as explicit entropy production rate in what follows, has been evaluated in various active dynamics, either particle-based~\cite{Speck2016, Nardini2016, Mandal2017, Shankar2018, Seifert2018, Bo2019} or field theories~\cite{Nardini2017, Ramaswamy2018, Murrell2019}, to assess unambiguously the deviation from equilibrium. Our approach differs in that we not only account for the irreversibility of active fields, but also for that of underlying chemical degrees of freedom. In this extended phase space, provided that it accounts for all the relevant hydrodynamic fields, the irreversibility indeed measures the heat dissipated by the {\it entire} system.

Following standard procedures~\cite{Onsager1953, Martin1973, Dominicis1975}, the dynamic action $\cal A$ which sets the path probability ${\cal P}\sim{\rm e}^{-{\cal A}}$ reads
\begin{equation}\label{eq:action}
	\begin{aligned}
		{\cal A} &= \frac{1}{4T} \int_0^t\int_V \bigg\{ \big[{\bf J},\dot n\big] + {\mathbb L} \bigg[ \nabla\frac{\delta\cal F}{\delta\phi}, -\Delta\mu\bigg] \bigg\}
		\\
		&\quad\times {\mathbb L}^{-1} \bigg\{ \big[{\bf J},\dot n\big] + {\mathbb L} \bigg[ \nabla\frac{\delta\cal F}{\delta\phi}, -\Delta\mu\bigg] \bigg\}^\intercal \,{\rm d}{\bf r}{\rm d}s ,
	\end{aligned}
\end{equation}
where $\int_V$ is a spatial integral over the whole volume $V$ of the system, and ${\mathbb L}^{-1}$ is the inverse of $\mathbb L$. We regard the currents $\{{\bf J},\dot n\}$ and forces $\{-\nabla(\delta{\cal F}/\delta\phi),\Delta\mu\}$ as odd and even under time reversal, respectively. The action for the time-reversed dynamics ${\cal A}^{\rm R}$ is then deduced readily from~\eqref{eq:action} by flipping the sign of $[{\bf J}, \dot n]$. Substituting ${\cal P}\sim{\rm e}^{-{\cal A}}$ and ${\cal P}^{\rm R}\sim{\rm e}^{-{\cal A}^{\rm R}}$ into~\eqref{eq:entropy}, the dissipation rate follows from straightforward algebra as (see Appendix~\ref{app:entropy})
\begin{equation}\label{eq:entropy_a}
	\dot{\cal Q} = \int_V \bigg\langle \bigg\langle \dot n \Delta\mu - {\bf J}\cdot \nabla\frac{\delta\cal F}{\delta\phi} \bigg\rangle \bigg\rangle_t\,{\rm d}{\bf r} ,
\end{equation}
where $\lim_{t\to\infty} \frac{1}{t}\int_0^t \cdot \equiv \langle \cdot \rangle_t$ is the steady-state time average. For ergodic systems, the two averages are the same and one may be omitted. Hereafter, we use $\langle \cdot \rangle$ to denote both averages. The expression~\eqref{eq:entropy_a} features the sum of products between thermodynamic forces and conjugate currents, analogously to the dissipation rate in LIT~\cite{Mazur, Basu2008, Prost2017, Julicher2018, Markovich2019b}: This confirms that we embed active field theories within a thermodynamically consistent framework. Note that the product is interpreted here and in what follows with Stratonovich convention.

Integrating by parts the second term in~\eqref{eq:entropy_a} and using $\dot\phi=-\nabla\cdot{\bf J}$, we get $ \int_V \langle {\bf J}\cdot\nabla(\delta{\cal F}/\delta\phi) \rangle {\rm d}{\bf r} = {\rm d}\langle{\cal F}\rangle/{\rm d}t$, which vanishes in steady state, yielding
\begin{equation}\label{eq:entropy_b}
	\dot{\cal Q} = \int_V \langle \dot n \Delta\mu\rangle \,{\rm d}{\bf r} .
\end{equation}
As a result, the steady-state heat rate $\dot{\cal Q}$ equals the rate of work injected by the nonequilibrium drive $\Delta\mu$ to sustain the dynamics away from equilibrium: This is equivalent to the first law of thermodynamics, as expected when the path probabilities include all thermodynamically relevant fields. The expression~\eqref{eq:entropy_b} would actually be the same if instead $\dot n$ was held constant and $\Delta\mu$ allowed to fluctuate. For an equilibrium dynamics where $\Delta\mu$ derives from the chemical free energy ${\cal F}_{\rm ch}$, ($\Delta\mu = - \delta{\cal F}_{\rm ch}/\delta n$), 
the heat rate rate vanishes in steady state ($\dot{\cal Q} = -{\rm d}\langle{\cal F}_{\rm ch}\rangle/{\rm d}t = 0$), as expected.

Substituting the chemical dynamics~\eqref{eq:dyn_chem_g} in~\eqref{eq:entropy_b}, we deduce
\begin{equation}\label{eq:entropy_c}
	\dot{\cal Q} = \gamma V\Delta\mu^2 - \Delta\mu \int_V \bigg\langle {\bf C}\cdot\nabla\frac{\delta\cal F}{\delta\phi} - T \,\chi({\bf C}) \bigg\rangle \,{\rm d}{\bf r} .
\end{equation}
Hence, the heat rate can be separated into (i)~a homogeneous contribution $\gamma V\Delta\mu^2$ corresponding to a background term independent of the fluctuations of the active and chemical fields $\{\phi,n\}$, and (ii)~a contribution determined only by the fluctuations of the active field $\phi$, namely independent of that of $n$. The existence of $n$ however, is crucial in determining the form of the heat rate. This becomes clear below when we compare the heat rate with the explicit EPR, in which the dynamics of $n$ are not accounted for, see Eq.~\eqref{eq:entropy_c_bis}. Note that fast-relaxing fields which are deliberately omitted in our hydrodynamic description can only contribute to heat rate through an additional background term. Interestingly, this homogeneous contribution is eliminated when considering the difference of heat rates at constant $\Delta\mu$, for instance by changing parameters of the free energy $\cal F$: The heat-rate difference then depends only on how the fluctuations of the active field $\phi$ vary with such parameters.


\subsection{Generalized field dynamics}\label{sec:gen}

To demonstrate that our framework is indeed relevant for a large class of active field theories, we now consider the coupled dynamics of a conserved scalar field $\phi$ and a polar field ${\bf p}$:
\begin{equation}\label{eq:dyn_field}
	\begin{aligned}
		\dot\phi &= - \nabla\cdot{\bf J} ,
		\\
		{\bf J} &= - \lambda_\phi\nabla\frac{\delta\cal F}{\delta\phi} + \Delta_\phi\,{\bf C}_\phi + T \,{\boldsymbol\nu}_\phi({\bf C}_\phi) + {\boldsymbol\Lambda}_\phi ,
		\\
		\dot{\bf p} &= - \lambda_p \frac{\delta\cal F}{\delta\bf p} + \Delta_p\,{\bf C}_p + T \,{\boldsymbol\nu}_p({\bf C}_p) + {\boldsymbol\Lambda}_p ,
		\\
	\end{aligned}
\end{equation}
where $\lambda_\Omega$ and $\Delta_\Omega$ are respectively the mobility and the constant driving coefficient for $\Omega\in\{\phi, p\}$, and ${\bf C}_\Omega$ depend on $\{\phi,{\bf p}\}$ and their gradients. The noise term ${\boldsymbol\Lambda}_\Omega$ is Gaussian with zero mean and correlations given by 
\begin{equation}\label{eq:lambda_g}
	\big\langle\Lambda_{\Omega,\alpha}({\bf r},t)\Lambda_{\Omega',\beta}({\bf r}',t')\big\rangle = 2\lambda_\Omega T \delta_{\alpha\beta}\delta_{\Omega\Omega'}\delta({\bf r}-{\bf r}')\delta(t-t') .
\end{equation}
In what follows, we assume that $\Delta_\phi$ and $\Delta_p$ are independent, so that each one of ${\boldsymbol\nu}_\Omega$ is only determined by the corresponding ${\bf C}_\Omega$. The dynamics~(\ref{eq:dyn_field}-\ref{eq:lambda_g}) typically describe the coarse-grained dynamics of polar agents, ranging from vibrated grains~\cite{Narayan2007, Dauchot2010} to bird flocks~\cite{Vicsek1995, Toner1995} and aligning bacteria~\cite{Yeomans2012, Goldstein2013}. In practice, the dissipation rate for systems featuring other types of order parameters, such as a nematic tensor~\cite{Ramaswamy2003, Ramaswamy2006, Bertin2013, Ngo2014} or a non-conserved scalar field~\cite{Czajkowski2018}, extends straightforwardly from the results detailed below for the specific dynamics~(\ref{eq:dyn_field}-\ref{eq:lambda_g}).
Note that in all these examples both $\phi$ and ${\bf p}$ are structural order parameters, and are therefore even under time-reversal.

The spurious drift terms $T{\boldsymbol\nu}_\Omega$ were not considered in previous work. In what follows, we address cases where the driving coefficients $\Delta_\Omega$ are either odd or even under time reversal, and we assume that even (odd) driving represents a chemical potential difference $\Delta_\Omega=\Delta\mu_\Omega$ (chemical current $\Delta_\Omega=\dot n_\Omega/\gamma_\Omega$). We show in Appendix~\ref{app:discretisation} that the expression for ${\boldsymbol\nu}_\Omega$ in terms of ${\bf C}_\Omega$ depends on the choice for the parity of $\Delta_\Omega$. Besides, we put forward explicit cases where ${\boldsymbol\nu}_\Omega$ vanishes for judicious choices of the spatial discretization of gradient terms in ${\bf C}_\Omega$.

With the assumption that the fields $\{\phi,{\bf p}\}$ are even under time-reversal, LIT enforces that the form of the chemical dynamics is identical for either choice $\Delta_\Omega = \Delta\mu_\Omega$ or $\Delta_\Omega = \dot n_\Omega/\gamma_\Omega$~\cite{Mazur, Basu2008, Prost2017, Julicher2018, Markovich2019b}:
\begin{equation}\label{eq:dyn_chem_b}
	\begin{aligned}
		\dot n_\phi &= \gamma_\phi\Delta\mu_\phi - {\bf C}_\phi\cdot\nabla\frac{\delta\cal F}{\delta\phi} + T \,\chi_\phi({\bf C}_\phi) + \xi_\phi ,
		\\
		\dot n_p &= \gamma_p\Delta\mu_p - {\bf C}_p\cdot\frac{\delta\cal F}{\delta\bf p} + T \,\chi_p({\bf C}_p) + \xi_p ,
	\end{aligned}
\end{equation}
where $\gamma_\Omega$ is the chemical mobility, and $\xi_\Omega$ is a zero-mean Gaussian noise with correlations
\begin{equation}\label{eq:xi_g}
	\big\langle\xi_\Omega({\bf r},t)\xi_{\Omega'}({\bf r}',t') \big\rangle = 2\gamma_\Omega T \delta_{\Omega\Omega'}\delta({\bf r}-{\bf r}')\delta(t-t') .
\end{equation}
The noises $\xi_\Omega$ and ${\boldsymbol\Lambda}_\Omega$ are correlated only if the driving is even ($\Delta_\Omega = \Delta\mu_\Omega$), in which case
\begin{equation}\label{eq:dyn_chem_b_g}
	\big\langle\Lambda_{\Omega,\alpha}({\bf r},t)\xi_{\Omega'}({\bf r}',t') \big\rangle = 2 T C_{\Omega,\alpha}({\bf r},t) \delta_{\Omega\Omega'}\delta({\bf r}-{\bf r}')\delta(t-t') .
\end{equation}
The expression of $\chi_\Omega$ follows from that of ${\bf C}_\Omega$, as detailed in Appendix~\ref{app:discretisation}, and it differs according to whether $\Delta_\Omega = \Delta\mu_\Omega$ or $\Delta_\Omega = \dot n_\Omega/\gamma_\Omega$. We stress that in both cases the realizations of the active fields $\{\phi, {\bf p}\}$ are independent of the chemical dynamics~\eqref{eq:dyn_chem_b}. An alternative formulation of the dynamics, not considered explicitly here, consists in taking fluctuating $\Delta_\Omega$ in~\eqref{eq:dyn_field} and setting the conjugated chemical degree of freedom constant. Within this formulation, the chemical dynamics affects directly the active field dynamics, but the results for the heat rate below will remain the same, namely they only depend on the driving mechanism and not on how it affects the active dynamics.

The steady-state heat rate is now defined by
\begin{equation}
	\dot{\cal Q} = T \left< \underset{t\to\infty}{\lim} \frac{1}{t} \ln \frac{{\cal P}\big[\{{\bf J}, \dot{\bf p}, \dot n_\phi, \dot n_p\}_0^t\big]}{{\cal P}^{\rm R}\big[\{{\bf J}, \dot{\bf p}, \dot n_\phi, \dot n_p\}_0^t\big]} \right>,
\end{equation}
and it can be obtained following a similar procedure as that in Sec.~\ref{sec:entropy}. It again differs from the explicit entropy production rate ${\cal S}$ given by
\begin{equation}
	{\cal S} = \left< \underset{t\to\infty}{\lim} \frac{1}{t} \ln \frac{{\cal P}\big[\{{\bf J}, \dot{\bf p}\}_0^t\big]}{{\cal P}^{\rm R}\big[\{{\bf J}, \dot{\bf p}\}_0^t\big]} \right> .
\end{equation}
Identifying the thermodynamic forces and their conjugated currents as $\{-\nabla(\delta{\cal F}/\delta\phi), \Delta\mu_\phi, -\delta{\cal F}/\delta{\bf p}, \Delta\mu_p\}$ and $\{{\bf J}, \dot n_\phi, \dot{\bf p}, \dot n_p\}$, respectively, we get
\begin{equation}\label{eq:entropy_d}
	\dot{\cal Q} = \sum_{\Omega\in\{\phi, p\}} \int_V \langle \dot n_\Omega \, \Delta\mu_\Omega \rangle \,{\rm d}{\bf r} .
\end{equation}
The expression~\eqref{eq:entropy_d} is then valid for either $\Delta_\Omega = \Delta\mu_\Omega$ or $\Delta_\Omega = \dot n_\Omega/\gamma_\Omega$. It extends to an arbitrary number of active fields, potentially including other types of order parameters such as nematic tensors, and it remains valid when each active field couples to several chemical fields, see Appendix~\ref{app:entropy}: For any of these cases, the heat rate actually follows directly from the dynamics of $\dot n_\Omega$.

Substituting the chemical dynamics~\eqref{eq:dyn_chem_b} in~\eqref{eq:entropy_d}, when $\Delta_\Omega$ is a force ($\Delta_\Omega = \Delta\mu_\Omega$), we get
\begin{equation}\label{eq:entropy_force}
\begin{aligned}
\dot{\cal Q} &= \gamma_\phi V\Delta\mu_\phi^2 - \Delta\mu_\phi \int_V\bigg\langle {\bf C}_\phi\cdot\nabla\frac{\delta\cal F}{\delta\phi} - T\,\chi_\phi({\bf C}_\phi) \bigg\rangle \,{\rm d}{\bf r}
\\
&\quad + \gamma_pV\Delta\mu_p^2 - \Delta\mu_p \int_V \bigg\langle {\bf C}_p\cdot\frac{\delta\cal F}{\delta\bf p} - T\,\chi_p({\bf C}_p) \bigg\rangle \,{\rm d}{\bf r} .
\end{aligned}
\end{equation}
When $\Delta_\Omega$ is a current ($\Delta_\Omega = \dot n_\Omega/\gamma_\Omega$), we get instead
\begin{equation}\label{eq:entropy_current}
\begin{aligned}
\dot{\cal Q} &= \frac{V\dot n_\phi^2}{\gamma_\phi} + \frac{\dot n_\phi}{\gamma_\phi} \int_V \bigg\langle {\bf C}_\phi\cdot\nabla\frac{\delta\cal F}{\delta\phi} - T\,\chi_\phi({\bf C}_\phi) \bigg\rangle \,{\rm d}{\bf r}
\\
&\quad + \frac{V\dot n_p^2}{\gamma_p} + \frac{\dot n_p}{\gamma_p} \int_V \bigg\langle {\bf C}_p\cdot\frac{\delta\cal F}{\delta\bf p} - T\,\chi_p({\bf C}_p) \bigg\rangle \,{\rm d}{\bf r} .
\end{aligned}
\end{equation}
In general, $\Delta_\phi$ and $\Delta_p$ need not have the same parity, so that the heat rate can be a combination of the forms given in~(\ref{eq:entropy_force}-\ref{eq:entropy_current}).


\section{Applications to illustrative field theories}\label{sec:app}

Before applying our generic theory to quantify the heat rate in specific models, we compare the heat rate~\eqref{eq:entropy_c} with a measure of deviation from equilibrium obtained in previous works~\cite{Nardini2017, Ramaswamy2018, Murrell2019}. Substituting in~\eqref{eq:entropy_c} the expression of $\nabla(\delta{\cal F}/\delta\phi)$ taken from the dynamics~\eqref{eq:dyn_phi} yields
\begin{equation}\label{eq:heat_phi_sd}
\begin{aligned}
\dot{\cal Q} &= T{\cal S} + \frac{\Delta\mu^2}{\lambda} \int_V \big(\lambda\gamma - \big\langle{\bf C}^2\big\rangle\big) \,{\rm d}{\bf r} \\
&+ T \Delta\mu \int_V \left< \chi({\bf C}) - \frac{1}{\lambda} {\bf C} \cdot {\boldsymbol\nu}({\bf C}) -\frac{1}{T\lambda}  {\bf C} \cdot {\boldsymbol \Lambda} \right>   \,{\rm d}{\bf r}  \, .
\end{aligned}
\end{equation}
Because previous works did not account for the spurious-drift terms, a proper comparison requires consideration of the special case in which $\{{\boldsymbol\nu},\chi\}=\{{\bf 0},0\}$ and $\int_V\langle{\bf C}\cdot{\boldsymbol\Lambda}\rangle{\rm d}{\bf r}=0$. These expressions depend on the spatial discretization scheme. In Appendix~\ref{app:discretisation} we provide a recipe for calculating these expressions and give examples in which they vanish. In such cases
\begin{equation}\label{eq:entropy_c_bis}
\begin{aligned}
\dot{\cal Q} &= T{\cal S} + \frac{\Delta\mu^2}{\lambda} \int_V \big(\lambda\gamma - \big\langle{\bf C}^2\big\rangle\big) \,{\rm d}{\bf r}  ,
\end{aligned}
\end{equation}
and the explicit entropy production rate ${\cal S}$, which was also computed in previous studies~\cite{Nardini2017, Ramaswamy2018, Murrell2019}, reads
\begin{equation}\label{eq:S_phi}
{\cal S} = \frac{\Delta\mu}{\lambda T} \int_V \big\langle{\bf J}\cdot{\bf C}\big\rangle \,{\rm d}{\bf r} .
\end{equation}
Therefore,~\eqref{eq:entropy_c_bis} provides a connection between the thermodynamic heat rate $\dot{\cal Q}$ and the explicit entropy production rate ${\cal S}$. From the semi-positivity of the Onsager matrix $\mathbb L$, which ensures $\det{\mathbb L} = \lambda\gamma - {\bf C}^2\geq0$, it then follows that $T{\cal S}$ is a lower bound to $\dot{\cal Q}$. The bound is saturated when $\bf J$ and $\dot n$ are proportional ($\det{\mathbb L} = 0$): In such a case, the fluctuations of $\dot n$ are slaved to that of $\bf J$, so that the irreversibility of the whole dynamics can be found from trajectories of $\bf J$ alone.

For the generalized field dynamics that includes the dynamics of both $\phi$ and ${\bf p}$, we again consider the case in which $\{{\boldsymbol\nu}_\Omega,\chi_\Omega\}=\{{\bf 0},0\}$ and $\int_V\langle{\bf C}_\Omega\cdot{\boldsymbol\Lambda}_\Omega\rangle{\rm d}{\bf r}=0$. Then, substituting in~\eqref{eq:entropy_force} the expression of $\{\nabla(\delta{\cal F}/\delta\phi),\delta{\cal F}/\delta{\bf p}\}$ taken from the dynamics~\eqref{eq:dyn_field} for $\Delta_\Omega = \Delta\mu_\Omega$ we get
\begin{equation}\label{eq:entropy_force1}
\begin{aligned}
\dot{\cal Q} = T{\cal S} \,+ &\int_V \Bigg[ \frac{\Delta\mu^2_\phi}{\lambda_\phi} \big(\lambda_\phi\gamma_\phi - \big\langle {\bf C}_\phi^2 \big\rangle\big)
\\
&\quad + \frac{\Delta\mu^2_p}{\lambda_p} \big( \lambda_p\gamma_p - \big\langle {\bf C}_p^2 \big\rangle \big) \Bigg] {\rm d}{\bf r} ,
\end{aligned}
\end{equation}
where
\begin{equation}
{\cal S} = \int_V \bigg[ \frac{\Delta\mu_\phi}{\lambda_\phi T} \,\big\langle {\bf J}\cdot{\bf C}_\phi\big\rangle + \frac{\Delta\mu_p}{\lambda_p T} \,\big\langle\dot{\bf p}\cdot{\bf C}_p \big\rangle \bigg] {\rm d}{\bf r} .
\end{equation}
This shows explicitly the difference between the heat rate $\dot{\cal Q}$ and the explicit entropy production rate ${\cal S}$, similarly to~(\ref{eq:entropy_c_bis}). For $\Delta_\Omega = \dot n_\Omega/\gamma_\Omega$, we have instead
\begin{equation}\label{eq:entropy_current1}
\dot{\cal Q} = V \bigg(\,\frac{\dot n_\phi^2}{\gamma_\phi} + \frac{\dot n_p^2}{\gamma_p}\,\bigg) + T{\cal S} ,
\end{equation}
in which case the heat rate differs from the explicit entropy production rate by a background term.


We are now in the position to apply our generic theory to two popular active field theories: (i) the dynamics of a conserved scalar field which reproduces active phase separation, and (ii) the coupled dynamics of a conserved scalar field and a non-conserved polar field that captures the behavior of motile deformable droplets. 

\subsection{Active phase separation}\label{sec:phase}

To illustrate how our framework can quantify the heat rate to sustain a phase separation away from equilibrium, we consider a popular active field theory for a conserved scalar field $\phi$ that is even under time-reversal, known as Active Model B~\cite{Wittkowski2014, Nardini2017}. Taking the coupling term as ${\bf C} = - \nabla(\nabla\phi)^2$ in~\eqref{eq:dyn_phi} recovers the dynamical equation of Active Model B whenever the spurious drift term $T\boldsymbol\nu$ vanishes. From symmetry arguments, this coupling term is the lowest order in gradients and in $\phi$ which cannot be integrated into a free energy~\cite{Speck2013, Tailleur2013, Wittkowski2014, Nardini2017}. A term of the form $(\nabla\phi)(\nabla^2\phi)$ is potentially present at same order as $\nabla(\nabla\phi)^2$~\cite{Nardini2017, Nardini2018, Rapp2019}, yet both terms are equivalent in one spatial dimension, as we consider below.

The spurious drift terms $\{T\boldsymbol\nu,T\chi\}$ appearing in dynamics~(\ref{eq:dyn_phi}-\ref{eq:dyn_chem}) vanish when choosing a specific spatial discretization, as shown in Appendix~\ref{app:discretisation}. Then, there is no need to actually modify the dynamical equation already used in~\cite{Wittkowski2014, Nardini2017} to embed Active Model B in a thermodynamically consistent framework. For a constant chemical potential difference $\Delta\mu$, the dynamics follows as
\begin{equation}\label{eq:dyn_amb}
	\begin{aligned}
		\dot\phi &= \partial_x \bigg[ \partial_x\frac{\delta\cal F}{\delta\phi} + \Delta\mu\,\partial_x(\partial_x\phi)^2 + \Lambda \bigg] ,
		\\
		\dot n &= \gamma\Delta\mu + \big[\partial_x(\partial_x\phi)^2\big] \partial_x\frac{\delta\cal F}{\delta\phi} + \xi ,
	\end{aligned}
\end{equation}
where we have set the mobility $\lambda=1$, and $\{\Lambda,\xi\}$ are zero-mean Gaussian white noises with correlations proportional to the temperature $T$, as given in~\eqref{eq:lambda} and~\eqref{eq:dyn_chem}. The free energy $\cal F$ captures a phase separation between dilute and dense regions:
\begin{equation}\label{eq:free_energy}
	{\cal F} = \int \bigg[ f(\phi) + \frac{\kappa}{2}(\partial_x\phi)^2 \bigg] {\rm d}x ,
		\quad
	f(\phi) = \frac{a}{2}\phi^2 + \frac{b}{4}\phi^4 .
\end{equation}
In what follows, most of our results are valid for a generic $f$, and the specific form~\eqref{eq:free_energy} is used for explicit evaluation only.

The corresponding heat rate, as given in~\eqref{eq:entropy_c}, reads
\begin{equation}\label{eq:entropy_amb}
	\dot{\cal Q} = \gamma V\Delta\mu^2 +  \int_V \dot q \,{\rm d}x ,
	\quad
	\dot q = \Delta\mu \,\bigg\langle\big[\partial_x(\partial_x\phi)^2\big]\partial_x\frac{\delta\cal F}{\delta\phi}\bigg\rangle .
\end{equation}
The heat rate quantifies the irreversibility of the whole dynamics based on trajectories of the active current $J$ and of the chemical current $\dot n$, see~\eqref{eq:entropy}. The heat rate profile $\dot q(x)$ depends on the details of the dynamics via the parameters of the free-energy $\cal F$, the driving coefficient $\Delta\mu$, and the temperature $T$ which controls the amplitude of fluctuations. For strong fluctuations, namely high temperature $T$, we expect the heat rate to be uniformly dissipated in the system, with only a weak dependence on the details of the density profile. Conversely, in the regime of small $T$, the local heat rate should reveal the salient features of the density profile which require energy to be sustained.

To explore the connection between density profile and heat rate, we then rely on a small-noise treatment of the dynamics. Given that~\eqref{eq:entropy_amb} is fully determined by the fluctuations of $\phi$, independently of that of $n$, we focus on the dynamics of $\phi$ alone. Expanding the density field as $\phi = \phi_0 + \sqrt{T}\phi_1 + T\phi_2 + {\cal O}(T^{3/2})$ and substituting this ansatz in~\eqref{eq:dyn_amb}, the leading order equation yields the deterministic mean-field dynamics:
\begin{equation}\label{eq:mf}
	\dot\phi_0 = \partial_x^2 \big[ D_0 + \Delta\mu\,(\partial_x\phi_0)^2 \big] ,
	\quad
	D_0 = f'_0 - \kappa \partial_x^2\phi_0 ,
\end{equation}
where $f^{(n)}_0 = {\rm d}^nf/{\rm d}\phi^n$ at $\phi=\phi_0$. Hence, $\phi_0$ relaxes to a steady-state profile which can either be uniform or comprising phase-separated domains depending on free-energy parameters in~\eqref{eq:free_energy}, the global density $(1/V)\int_V\phi(x){\rm d}x$ and the driving parameter $\Delta\mu$~\cite{Nardini2017, Nardini2018}. At higher orders, $\phi_1$ and $\phi_2$ follow a set of coupled stochastic dynamics given by
\begin{equation}\label{eq:amb_exp}
	\begin{aligned}
		\dot\phi_1 &= \partial_x^2\big[ D_1 + 2\Delta\mu\,(\partial_x\phi_1)(\partial_x\phi_0) \big] + \partial_x \Lambda_0 ,
		\\
		\dot\phi_2 &= \partial_x^2\big\{ D_2 + \Delta\mu\,\big[2(\partial_x\phi_2)(\partial_x\phi_0) + (\partial_x\phi_1)^2 \big] \big\} ,
	\end{aligned}
\end{equation}
where
\begin{equation}\label{eq:amb_exp_b}
	D_1 = \big(f''_0- \kappa \partial_x^2\big) \phi_1 ,
	\quad
	D_2 = \big(f''_0- \kappa \partial_x^2\big) \phi_2 + f'''_0\phi_1^2/2 ,
\end{equation}
and $\Lambda_0$ is a zero-mean Gaussian white noise with correlations $\langle\Lambda_0(x,t)\Lambda_0(x',t')\rangle=2\delta(x-x')\delta(t-t')$. Owing to the linearity of $D_1$ in~(\ref{eq:amb_exp}-\ref{eq:amb_exp_b}), $\phi_1$ has Gaussian fluctuations. Substituting the density ansatz in~\eqref{eq:entropy_amb}, we get
\begin{equation}\label{eq:entropy_amb_exp}
	\dot q = \varepsilon_0 + T\varepsilon_1 + {\cal O}(T^2) ,
\end{equation}
where
\begin{equation}\label{eq:entropy_amb_exp_b}
	\begin{aligned}
		\varepsilon_0 = - \Delta\mu&\, (\partial_x\phi_0)^2\partial_x^2D_0 ,
		\\
		\varepsilon_1 = - \Delta\mu&\, \Big[\,\big\langle(\partial_x\phi_1)^2\big\rangle\partial_x^2D_0 + 2 (\partial_x\phi_0)\big\langle (\partial_x\phi_1)\partial_x^2D_1 \big\rangle
		\\
		& + (\partial_x\phi_0)^2 \big\langle\partial_x^2D_2 \big\rangle + 2 (\partial_x\phi_0)\big\langle\partial_x\phi_2\big\rangle \partial_x^2D_0 \,\Big] .
	\end{aligned}
\end{equation}
The expressions~(\ref{eq:entropy_amb_exp}-\ref{eq:entropy_amb_exp_b}) give the leading orders of heat rate at small noise for an arbitrary $f$.

\begin{figure}
	\centering
	\includegraphics[width=\linewidth]{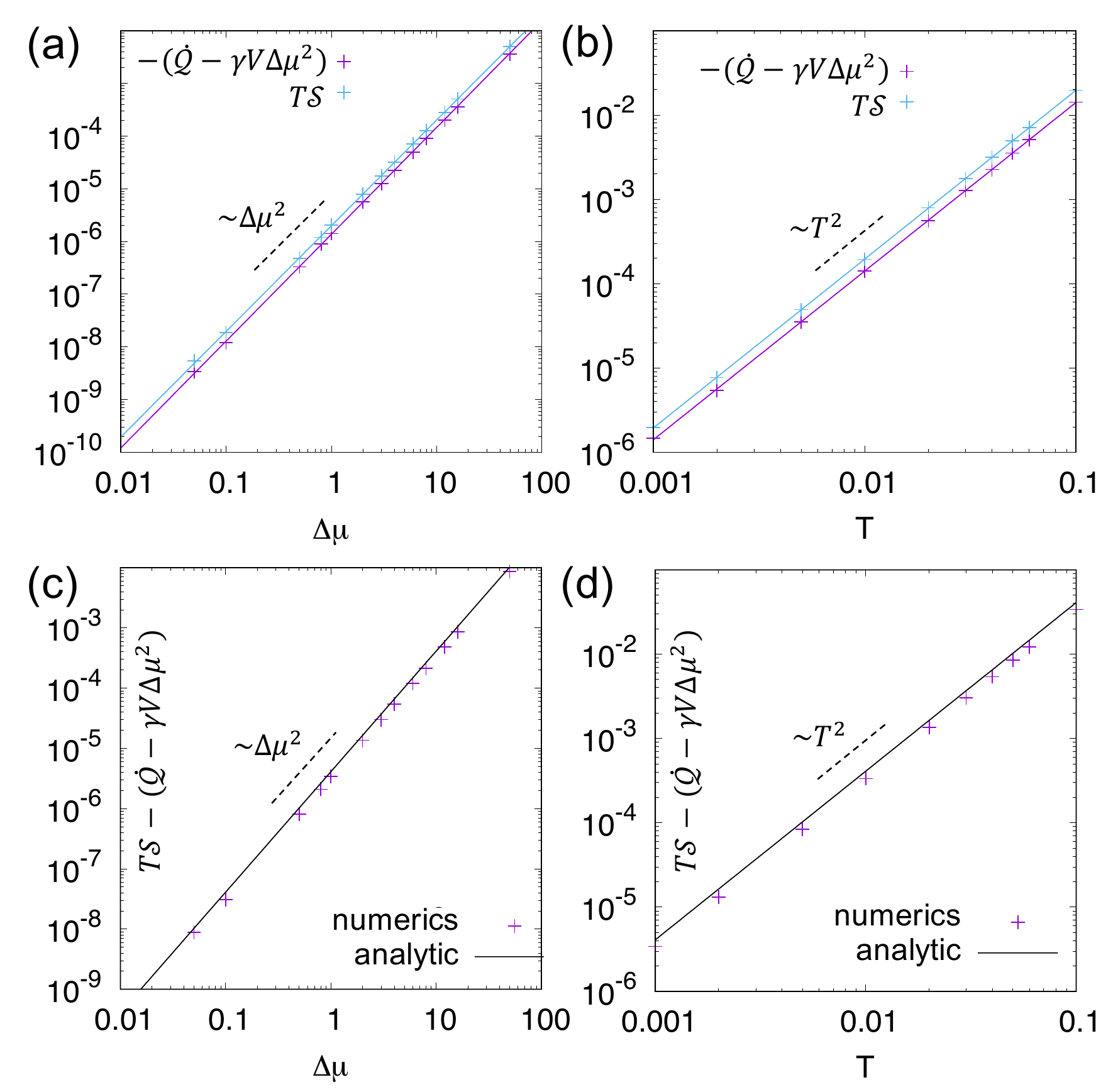}
	\caption{\label{fig2}
		In the absence of phase separation, namely for a homogeneous average profile of density ($\langle\phi(x)\rangle = {\rm cst}$), the non-trivial contribution to heat rate $\dot{\cal Q}-\gamma V\Delta\mu^2$ and the density field irreversibility $T{\cal S}$ scale like $T^2$ at small temperature $T$, and as $\Delta\mu^2$ at small driving parameter $\Delta\mu$, as shown respectively in (a,b) where solid lines are guide lines. Their difference also exhibits similar scalings in these regimes, as shown in (c,d), and it is in good agreement with our prediction~\eqref{eq:heat_diff} reported in black solid lines.
		Simulation details in Appendix~\ref{app:simulations}.
		Parameters: $-a=b=0.25$, $\kappa=4$, $\bar\phi=1$, $V=128$, (a,c)~$T=10^{-3}$, (b,d)~$\Delta\mu=1$.
	}
\end{figure}

For a homogeneous profile ($\phi_0(x)={\rm cst}$), the leading and first orders of the small noise expansion~(\ref{eq:entropy_amb_exp}-\ref{eq:entropy_amb_exp_b}) vanish ($\varepsilon_0=0$ and $\varepsilon_1=0$). Then, the non-trivial contribution to heat rate $\dot{\cal Q}-\gamma V\Delta\mu^2$ scales like $T^2$ at small $T$, and it also behaves like $\Delta\mu^2$ at small $\Delta\mu$, as confirmed by the numerical results in Figs.~\ref{fig2}(a-b). Therefore, in the absence of any density pattern, one can make $\dot{\cal Q}-\gamma V\Delta\mu^2$ arbitrarily small by reducing either the amplitude of fluctuations $T$ or the driving parameter $\Delta\mu$. In particular, at vanishing $T$, the uniform density profile is identical to that of Passive Model B, namely for $\Delta\mu=0$, which explains why $\dot{\cal Q}-\gamma V\Delta\mu^2$ also vanishes.

Previous works quantified irreversibility based on trajectories of the active current $J$ only~\cite{Nardini2017, Ramaswamy2018}, without considering that of the chemical current $\dot n$:
\begin{equation}\label{eq:entropy_phi_amb}
	T{\cal S} = \int_V \sigma \,{\rm d}x ,
	\quad
 	\sigma = - \Delta\mu \,\big\langle J \partial_x(\partial_x \phi)^2\big\rangle ,
\end{equation}
as defined in~\eqref{eq:S_phi}. The irreversibility measure $T\cal S$ has similar scalings as that of $\dot{\cal Q}-\gamma V\Delta\mu^2$ at small $T$ and small $\Delta\mu$, as shown in Figs.~\ref{fig2}(a-b). Interestingly, while the heat rate $\dot{\cal Q}$ converges to the finite value $\gamma V\Delta\mu^2$ at zero $T$, the irreversibility measure $T{\cal S}$ vanishes in this limit: The former captures the consumption of underlying chemicals, whereas the latter only sees an effective equilibrium dynamics. Moreover, the difference between $T\cal S$ and $\dot{\cal Q}-\gamma V\Delta\mu^2$ is $\Delta\mu^2\int_V \langle[\partial_x(\partial_x\phi)^2]^2\rangle{\rm d}x$ according to~\eqref{eq:entropy_c_bis}. We compute analytically this difference in Appendix~\ref{app:entropy} to show that it also scales like $T^2$ and $\Delta\mu^2$ at small $T$ and small $\Delta\mu$, respectively, as confirmed by our numerics in Figs.~\ref{fig2}(c,d).

\begin{figure}
	\centering
	\includegraphics[width=\linewidth]{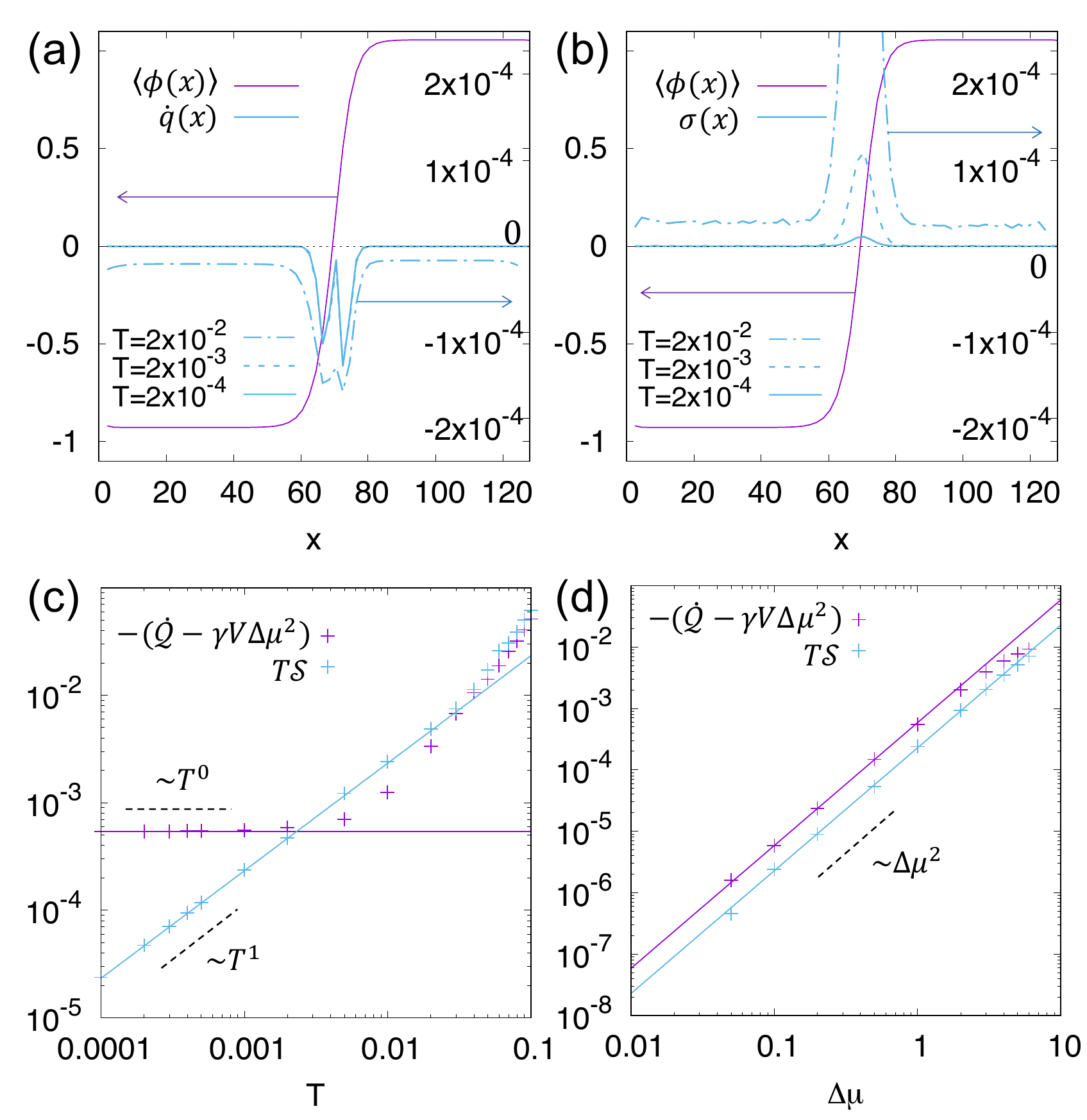}
	\caption{\label{fig3}
		(a-b)~The average profile of density $\langle\phi(x)\rangle$ shows a separation between dilute ($\langle\phi(x)\rangle<0$) and dense ($\langle\phi(x)\rangle>0$) phases. The corresponding profiles of heat rate $\dot q(x)$ and of irreversibility measure $\sigma(x)$, given respectively in~\eqref{eq:entropy_amb} and~\eqref{eq:entropy_phi_amb}, are flat in bulk regions and vary rapidly across the interface.
		(c)~The non-trivial contribution to heat rate $\dot{\cal Q}-\gamma V\Delta\mu^2 = \int_V \dot q {\rm d}x$ reaches a finite value at $T=0$, whereas the irreversibility measure $T{\cal S} = \int_V \sigma {\rm d}x$ vanishes.
		(d)~$\dot{\cal Q}-\gamma V\Delta\mu^2$ and $T{\cal S}$ respectively increase and decrease with the driving parameter $\Delta\mu$, and both scale as $\Delta\mu^2$.
		Simulation details in Appendix~\ref{app:simulations}.
		Parameters: $-a=b=0.25$, $\kappa=4$, $\bar\phi=0$, $V=128$, (a,b)~$\{\Delta\mu,T\}=\{2,10^{-2}\}$, (c)~$\Delta\mu=1$, (d)~$T=10^{-3}$.
	}
\end{figure}

For a phase-separated profile ($\phi_0(x)\neq{\rm cst}$), as shown in Fig.~\ref{fig3}(a-b), the leading order of $\dot{\cal Q}-\gamma V\Delta\mu^2$ scales like $T^0$, since now $\varepsilon_0\neq0$, and it reaches a finite value at $T=0$. Hence, the heat rate $\dot{\cal Q}$ is not only determined by the background term $\gamma V\Delta\mu^2$ at zero temperature, it now also depends on the mean-field density profile. In contrast, $T{\cal S}$ scales like $T$ and thus vanishes at $T=0$, see Fig.~\ref{fig3}(c), as already reported in~\cite{Nardini2017}. The different scalings of $\dot{\cal Q}$ and $T\cal S$ in this regime reveal that the former is affected by the existence of two separated phases, whereas the latter does not allow one to distinguish the active phase separation from its passive counterpart.  This clearly illustrates that the irreversibility shown by the active current $J$ alone, when the underlying chemical flux $\dot n$ is not monitored, cannot capture the full energetic cost of creating phase separation away from equilibrium. In other words, if one were to propose $T{\cal S}$ as a thermodynamically consistent measure of the full energetic cost, based on the explicit entropy production rate $\cal S$ which discards the driving field fluctuations, then a nonequilibrium phase separation could be sustained at zero cost, in contradiction with the basics of thermodynamics.

The heat profile $\dot q(x)$ given in~\eqref{eq:entropy_amb} not only provides information about where heat is dissipated, it also quantifies how the average chemical current $\langle\dot n(x)\rangle$ varies in space. At small temperature, it is constant in the dense and dilute phases, where the density profile is flat, and it has a non-monotonic behavior across the interface, as shown in Fig.~\ref{fig3}(a): The system dissipates less heat at the interface than in the bulk, and it does so by reducing locally the chemical current to accommodate for density gradients. Likewise, the profile $\sigma(x)$ in~\eqref{eq:entropy_phi_amb} is flat in the bulk and varies strongly at the interface~\cite{Nardini2017}. Yet, now both the bulk and interface contributions vanish at zero temperature, see Fig.~\ref{fig3}(b), consistently with the fact that $T{\cal S}$ vanishes in this regime. Moreover, both $T{\cal S}$ and $\dot{\cal Q}-\gamma V\Delta\mu^2$ scale as $\Delta\mu^2$ for small $\Delta\mu$, as shown in Fig.~\ref{fig3}(d), similarly to the case of a homogeneous density profile: The energetic cost $\dot{\cal Q}$ and the irreversibility measure $T\cal S$ vanish at zero $\Delta\mu$, since Active Model B becomes Passive Model B in this regime.


\subsection{Motile polar droplets}\label{sec:drop}

As a second example of how our formalism may be used, we now turn to study the coupled dynamics of a polar field ${\bf p}$ and a scalar field $\phi$. In our case, they represent the local polarization and density of active components, respectively, for instance self-propelled particles with aligning interactions~\cite{Vicsek1995, Narayan2007, Dauchot2010}, and are even under time-reversal. Our aim is to capture the emergence of complex order beyond the case of a phase separation, already discussed in Sec.~\ref{sec:phase}, by incorporating the possibility of observing a nonequilibrium polar order~\cite{Markovich2019,Markovich2019b}. We focus on dynamics with one spatial dimension for simplicity. A minimal ingredient to allow for a nonequilibrium advection of the fields then consists in taking the coupling terms as $C_\phi = \phi p$ and $C_p = - p \partial_x p$ in~\eqref{eq:dyn_field}. When the spurious drift terms $\{\nu_\phi, \nu_p\}$ vanish, one recovers the dynamical equations used to model actomyosin droplets in the absence of hydrodynamic flow, {\it e.g.} due to substrate friction, as detailed in~\cite{Tjhung2012} for instance. In general, such type of coupling terms appears naturally when coarse-graining the dynamics of aligning active agents~\cite{Bertin2009, Farrell2012}, and they also follow from symmetry arguments~\cite{Toner1995}.

To show that our framework is also applicable for odd driving, we now choose to treat constant chemical currents, namely $\Delta_\phi=\dot n_\phi/\gamma_\phi$ and $\Delta_p=\dot n_p/\gamma_p$. The spurious drift terms $\{\nu_\phi, \chi_\phi, \nu_p, \chi_p\}$ in~(\ref{eq:dyn_field}-\ref{eq:dyn_chem_b_g}) all vanish when choosing an appropriate spatial discretization, as detailed in Appendix~\ref{app:discretisation}.  The dynamics are then given by
\begin{equation}\label{eq:dyn_motile_a}
	\begin{aligned}
		\dot\phi &= \partial_x \bigg( \partial_x\frac{\delta\cal F}{\delta\phi} - \dot n_\phi \,\phi p + \Lambda_\phi \bigg) ,
		\\
		\dot n_\phi &= \Delta\mu_\phi - \phi p \,\partial_x\frac{\delta\cal F}{\delta\phi} + \xi_\phi ,
	\end{aligned}
\end{equation}
and
\begin{equation}\label{eq:dyn_motile_b}
	\begin{aligned}
		\dot p &= - \frac{\delta\cal F}{\delta p} - \dot n_p \,p\partial_xp + \Lambda_p ,
		\\
		\dot n_p &= \Delta\mu_p + p (\partial_x p) \frac{\delta\cal F}{\delta p} + \xi_p .
	\end{aligned}
\end{equation}
We have set the mobilities $\{\lambda_\phi, \lambda_p, \gamma_\phi, \gamma_p\}$ all equal to $1$, and $\{\Lambda_\phi,\Lambda_p,\xi_\phi,\xi_p\}$ are zero-mean Gaussian white noises with correlations proportional to $T$, as given in~\eqref{eq:lambda_g} and~(\ref{eq:xi_g}-\ref{eq:dyn_chem_b_g}). Inspired by recent works~\cite{Tjhung2012, Tjhung2015}, we take the free energy $\cal F$ which leads to the formation of motile and quiescent regions:
\begin{equation}\label{eq:free_energy_motile}
	\begin{aligned}
		{\cal F} &= \int \bigg[ f(\phi, p) + \frac{\kappa}{2}(\partial_x\phi)^2 + \frac{K}{2}(\partial_x p)^2 \bigg] {\rm d}x ,
		\\
		f(\phi, p) &= \frac{a}{4} \phi^2(\phi-2\bar\phi)^2 + \frac{A}{4} p^2 \big[p^2 +2(\bar\phi -\phi) \big] ,
	\end{aligned}
\end{equation}
where the coefficients $\{a, \bar\phi, A, K, \kappa\}$ are all positive. At equilibrium ($\dot n_\phi=0$ and $\dot n_p=0$), the system undergoes a phase separation whenever the global density $(1/V)\int_V\phi(x){\rm d}x$ is positive and less than $\phi_{\rm d} = \bar\phi\big[1+\sqrt{1+A/(2a\bar\phi^2)}\big]$, yielding coexistence between the dilute isotropic phase $\{\phi,p\}=\{0,0\}$ and the dense polar phase $\{\phi,p\}=\{\phi_{\rm d},\pm\sqrt{\phi_{\rm d}-\bar\phi}\}$.

The associated heat rate~\eqref{eq:entropy_current} reads
\begin{equation}\label{eq:entropy_motile}
	\begin{aligned}
		\dot{\cal Q} &= V(\dot n_\phi^2 + \dot n_p^2) + \int_V \dot q \,{\rm d}x ,
		\\
		\dot q &=  \dot n_\phi \,\bigg\langle \phi p \,\partial_x\frac{\delta\cal F}{\delta\phi} \bigg\rangle - \dot n_p \,\bigg\langle p(\partial_x p)\,\frac{\delta\cal F}{\delta p} \bigg\rangle .
	\end{aligned}
\end{equation}
To explore how the heat rate behaves at small temperature $T$, we again expand the fields as $\phi = \phi_0 + \sqrt{T}\phi_1 + {\cal O}(T)$ and $p = p_0 + \sqrt{T}p_1 + {\cal O}(T)$. The mean-field dynamics follows from~(\ref{eq:dyn_motile_a}-\ref{eq:dyn_motile_b}) as
\begin{equation}
	\begin{aligned}
		\dot\phi_0 + \dot n_\phi \,\partial_x(\phi_0 p_0) &= \partial_x^2 D_{\phi,0} ,
		\quad
		D_{\phi,0} = f_\phi - \kappa\nabla^2\phi_0 ,
		\\
		\dot p_0 + \dot n_p \,p_0 \partial_x p_0 &= - D_{p,0} ,
		\quad\;
		D_{p,0} = f_p - K\nabla^2 p_0 ,
	\end{aligned}
\end{equation}
where $f_\Omega = \partial_\Omega f(\phi_0,p_0)$ for $\Omega\in\{\phi,p\}$. The first correction to the mean-field profile reads
\begin{equation}\label{eq:motile_exp}
	\begin{aligned}
		\dot\phi_1 + \dot n_\phi \,\partial_x(\phi_0 p_1 + \phi_1 p_0) &= \partial_x^2 D_{\phi,1} + \partial_x \Lambda_{\phi,0} ,
		\\
		\dot p_1 + \dot n_p \,(p_0\partial_x p_1 + p_1 \partial_x p_0) &= - D_{p,1} + \Lambda_{p,0} ,
	\end{aligned}
\end{equation}
in terms of
\begin{equation}\label{eq:motile_exp_b}
	\begin{aligned}
		D_{\phi,1} &= \big(f_{\phi\phi} - \kappa \partial_x^2\big) \phi_1 + f_{\phi p} p_1 ,
		\\
		D_{p,1} &= \big(f_{pp} - K\partial_x^2\big) p_1 + f_{\phi p}\phi_1 ,
	\end{aligned}
\end{equation}
where $\Lambda_{\Omega,0}$ are zero-mean Gaussian white noises with correlations $\langle\Lambda_{\Omega,0}(x,t)\Lambda_{\Omega',0}(x',t') \rangle = 2 \delta_{\Omega\Omega'}\delta(x-x')\delta(t-t')$. As for the expansion in Sec.~\ref{sec:phase}, the active fields at first order $\{\phi_1, p_1\}$ have Gaussian statistics. The heat rate~\eqref{eq:entropy_motile} can then be expanded in the form~\eqref{eq:entropy_amb_exp}, where $\{\varepsilon_0, \varepsilon_1\}$ now read
\begin{equation}\label{eq:entropy_motile_exp_b}
	\begin{aligned}
		\varepsilon_0 &= \dot n_\phi\, \phi_0 p_0 \,\partial_x D_{\phi,0} - \dot n_p \,p_0(\partial_x p_0) D_{p,0} ,
		\\
		\varepsilon_1 &= \dot n_\phi \,\Big[ \,\big\langle\phi_1 p_1\big\rangle \,\partial_x D_{\phi,0} + \big\langle(\phi_1 p_0 + \phi_0 p_1) \,\partial_x D_{\phi,1}\big\rangle \,\Big]
		\\
		&\quad - \dot n_p \,\Big[ \,\big\langle p_1\partial_x p_1\big\rangle D_{p,0} + \big\langle (p_1\partial_x p_0 + p_0\partial_x p_1) D_{p,1} \big\rangle \,\Big] .
	\end{aligned}
\end{equation}
As a result, \eqref{eq:entropy_motile_exp_b} provides the leading orders of heat rate at small temperature for a given free-energy density $f$.

\begin{figure}
	\centering
	\includegraphics[width=.7\linewidth]{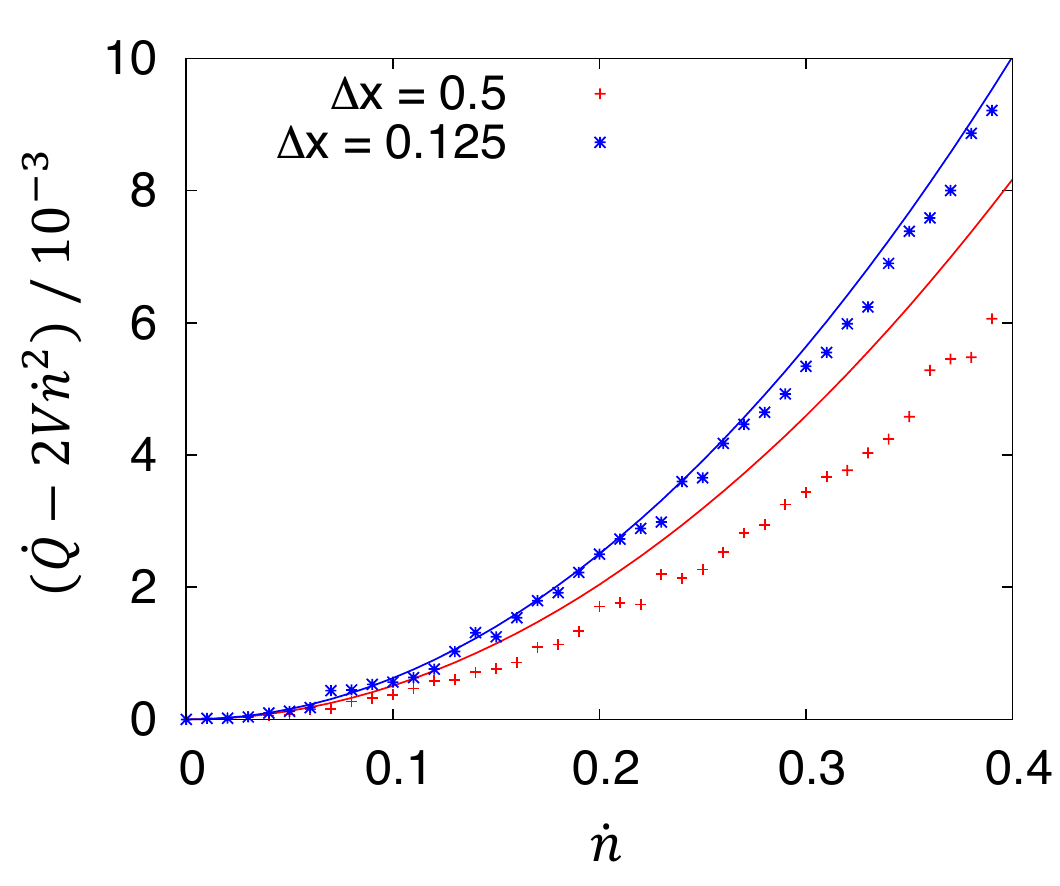}
	\caption{\label{fig4}
		In the absence of a droplet, namely for homogeneous average profiles of density ($\langle\phi(x)\rangle={\rm cst}$) and polarization ($\langle p(x)\rangle={\rm cst}$), the nontrivial contribution to heat rate $\dot{\cal Q}-2V\dot n^2$ scales like $T^2$ at small temperature $T$, and as $\dot n^2$ at small driving parameter $\dot n_\phi=\dot n_p\equiv\dot n$. The numerical measurements get closer to our analytical predictions~\eqref{eq:spectrum}, shown respectively in markers and solid lines, when the lattice spacing $\Delta x$ decreases, as expected.
		Simulation details in Appendix~\ref{app:simulations}.
		Parameters: $a=A=\kappa=K=\bar\phi=1$, $T=10^{-3}$, $V=64$.
	}
\end{figure}

In the homogeneous state ($\phi_0(x)={\rm cst}$ and $p_0(x)={\rm cst}$), the mean-field contribution to $\dot{\cal Q}-V(\dot n_\phi^2 + \dot n_p^2)$ vanishes ($\varepsilon_0=0$), yet the first order correction provides a non-zero contribution ($\varepsilon_1\neq0$): The non-trivial contribution to heat rate $\dot{\cal Q}-V(\dot n_\phi^2 + \dot n_p^2)$ scales like $T$, in line with what was found in~\cite{Ramaswamy2018} and in contrast with the $T^2$ scaling for the conserved dynamics of the scalar field $\phi$ in Sec.~\ref{sec:phase}. We compute analytically this contribution in terms of the dynamical parameters, as detailed in Appendix~\ref{app:entropy}. For simplicity, we choose the driving parameters in the dynamics of $\phi$ and $p$ to be equal ($\dot n_\phi = \dot n_p\equiv\dot n$), in which case $\dot{\cal Q}-2V\dot n^2$ behaves as $\dot n^2$: This scaling is confirmed by our numerical results in Fig.~\ref{fig4}. Note that our analytical result (see Eq.~\eqref{eq:spectrum}) depends on the lattice spacing through an ultra-violet cutoff.

\begin{figure}
	\centering
	\includegraphics[width=\linewidth]{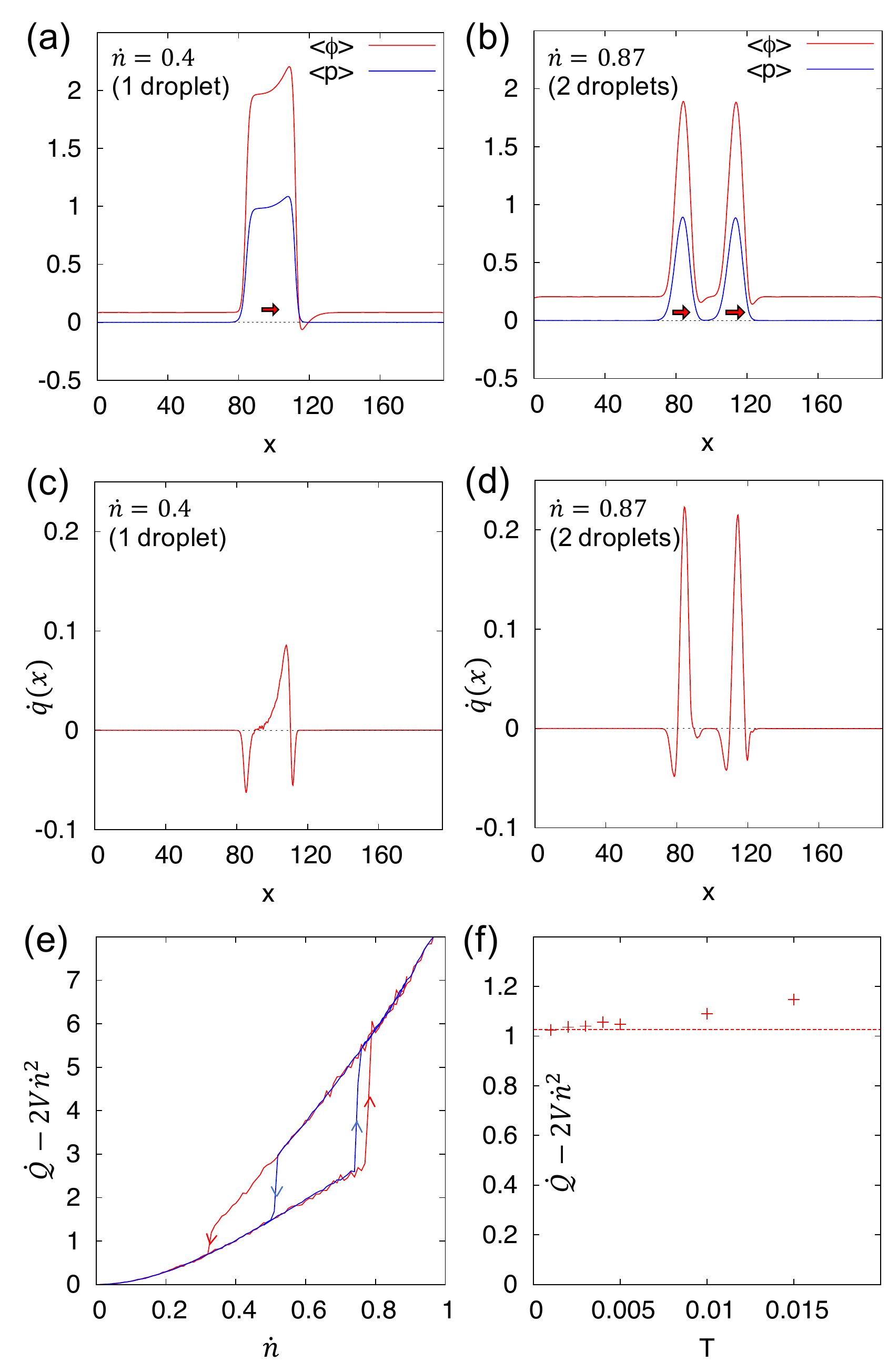}
	\caption{\label{fig5}
		(a-b)~The average profiles of density $\langle\phi(x)\rangle$ and polarization $\langle p(x)\rangle$, reported in the co-moving frame of droplets moving towards $x>0$, show that polarization is non-zero only within droplets and its profile has a front-tail asymmetry. For multiple droplets, each one of them moves at same velocity with a fixed separation distance.
		(c-d)~The corresponding average profiles of heat rate $\dot q(x)$, given in~\eqref{eq:entropy_motile}, are negative at the interface and increase monotonically from tail to front of each droplet.
		(e)~The nontrivial contribution to heat rate $\dot{\cal Q}-2V\dot n^2 = \int_V \dot q {\rm d}x$ increases with the driving parameter $\dot n_\phi=\dot n_p\equiv\dot n$ of the motile polar droplets. Above a critical value of $\dot n$, when the droplet splits into two droplets, we observe a discontinuity of $\dot{\cal Q}-2V\dot n^2$ associated with a hysteresis loop whose area increases with the speed at which $\dot n$ varies.
		(f)~$\dot{\cal Q}-2V\dot n^2$ reaches a finite value at $T=0$.
		Simulation details in Appendix~\ref{app:simulations}.
		Parameters: $a=A=\kappa=K=\bar\phi=1$, $V=196$, (a-e)~$T=10^{-3}$, (f)~$\dot n=0.4$.
	}
\end{figure}

For a droplet state ($\phi_0(x)\neq{\rm cst}$ and $p_0(x)\neq{\rm cst}$), as shown in Fig.~\ref{fig5}(a), $\dot{\cal Q}-2V\dot n^2$ scales like $T^0$ since the leading order is now determined by $\varepsilon_0\neq 0$. Analogously to the phase separated state in Sec.~\ref{sec:phase}, such a scaling implies that the heat rate $\dot{\cal Q}$ depends on the details of the density and polarization profiles even at vanishing temperature. Increasing the value of $\dot n$ splits the droplet into several ones which move in the same direction with a fixed separating distance, as shown in Fig.~\ref{fig5}(b). For one droplet, the heat profile $\dot q(x)$ in~\eqref{eq:entropy_motile} is peaked with negative values at the droplet interface, and it increases continuously with positive values from tail to head, see Fig.~\ref{fig5}(c). This behavior is qualitatively similar for two droplets, see Fig.~\ref{fig5}(d). In contrast with the case of a purely scalar field theory in Sec.~\ref{sec:phase}, the heat profile is now non-zero not only at the interface, but also in the dense phase: This stems from the density and polarization profiles of droplets being non-flat. Moreover, the fact that $\dot q(x)$ can have both signs illustrates that the local heat rate can be either above or below the background dissipation $2\dot n^2$. Therefore $\dot q(x)-2\dot n^2$ can potentially be negative {\it locally}, as long as the overall heat rate $\dot{\cal Q}$ stays positive, which corresponds to extracting energy from the thermostat at specific locations.

Interestingly, $\dot{\cal Q}-2V\dot n^2$ as a function of $\dot n$ displays a discontinuity when the number of droplets varies, see transition between one and two droplets reported in Fig.~\ref{fig5}(e). This shows that the total heat rate is strongly affected by the transition between different patterns, hence it can potentially be regarded as a relevant observable to characterize transitions, in line with previous results in particle-based active dynamics~\cite{Shim2016, Spinney2018}. Varying $\dot n$ linearly in time, we observe a hysteretic behavior so that the area of the loop in $\dot{\cal Q}-2V\dot n^2$ vs $\dot n$ space increases with the driving parameter velocity ${\rm d} \dot n/{\rm d}t$. Moreover, Fig.~\ref{fig5}(f) confirms that $\dot{\cal Q}-2V\dot n^2$ scales like $T^0$ at small noise, as predicted analytically.


\section{Conclusion}\label{sec:conclusion}

Building the thermodynamics of active matter is a major challenge of modern nonequilibrium statistical mechanics. By combining first-principles and phenomenological arguments, it aims at quantifying and predicting anomalous properties in terms of a few well-chosen observables~\cite{Takatori2015, Solon2015b, Shim2016, Nemoto2018, Paliwal2018, Spinney2018, Guioth2019, Suri2019}. Following this route, the irreversibility of active dynamics has recently attracted much attention, since it provides an unambiguous measure of the distance from equilibrium: It is quantified by the explicit entropy production rate (EPR) which compares forward and time-reversed realizations of the dynamics~\cite{Speck2016, Nardini2016, Nardini2017, Mandal2017, Maggi2017, Shankar2018, Seifert2018, Ramaswamy2018, Bo2019, Murrell2019}.

At microscopic level, the particle-based EPR can be related to the amount of heat dissipated by the system, though this relation can be more intricate for active systems~\cite{Bo2019} than for thermal ones~\cite{Sekimoto1998, Seifert2012}. At the hydrodynamic level, the connection between heat and explicit EPR is generally lost, so that the physical motivation for evaluating the explicit EPR in active field theories is sometimes unclear. Indeed, the heat rate is proportional to the total EPR provided that the latter measures the irreversibility of all hydrodynamic fields~\cite{Sekimoto1998, Seifert2012}. In contrast, the explicit EPR, which focuses on the irreversibility of active fields alone and discards the fluctuations of underlying driving fields, captures only a partial contribution to the heat rate. In practice, evaluating the total heat rate is then a challenge of modeling properly the coupling between active and driving fields.

In this paper, we have shown that the heat rate can be decomposed into a background contribution, independent of the active field, and a non-trivial contribution that dictates how the emerging order affects the energy cost. Importantly, the latter can be deduced systematically from the active field dynamics alone, provided that the equations of motion for active and driving fields are thermodynamically consistent, namely that the connection to surrounding thermostat is properly taken into account~\cite{Sekimoto1998, Seifert2012}. To ensure such a connection, we have embedded active field theories within linear irreversible thermodynamics~\cite{Mazur}, inspired by previous works~\cite{Kruse2004, Joanny2009, Prost2017, Ramaswamy2017, Ramaswamy2018}. It amounts to considering underlying degrees of freedom as the basis of the nonequilibrium drive of the dynamics. Yet, at variance with previous studies~\cite{Kruse2004, Joanny2009, Prost2017, Ramaswamy2017, Ramaswamy2018}, we now consider explicitly the fluctuations of these driving fields.

Thermodynamic consistency enforces some spurious drift terms in the dynamics that are proportional to noise amplitude. As such, our framework is distinct from the approach commonly followed to derive active field theories, based either on symmetry arguments~\cite{Toner1995, Wittkowski2014, Nardini2017, Nardini2018, Rapp2019, Markovich2019, Markovich2019b} or explicit coarse-graining of microscopic dynamics~\cite{Bertin2009, Farrell2012, Speck2013, Tailleur2013, Markovich2020}, since it enforces dynamical terms often neglected in these theories. In practice, while being conceptually important, the spurious drift terms can be made to vanish by judicious choices of spatial discretization. More generally, these terms do not affect the mean-field behavior of the system at vanishing noise, so that the emergent dynamics and structure are still consistent with the existing literature of active field theory in this regime.

Within our framework, the dynamics of the active fields can be read out independently of that of driving fields, so that the latter may be regarded as {\it hidden} degrees of freedom. Moreover, the spatial decomposition of heat rate can be evaluated in terms of active fields only. This supports the fact that the emerging dynamics and patterns of active fields alone provide direct access to spatial variations of heat rate, without need to measure the fluctuations of hidden degrees of freedom. Note that fast-relaxing variables are neglected by our hydrodynamic description, which potentially provide additional contributions to the heat rate. Yet, provided that there is indeed a clear time scale separation between the fluctuations of hydrodynamic fields, either active or driving fields, and that of other neglected variables, any additional contribution to heat rate only changes the constant background term: Thus, it does not affect the connection between active field patterns and spatial variations of heat rate.

To demonstrate the practical relevance of our approach, we have evaluated the heat rate in two popular active field theories: (i)~the dynamics of a conserved scalar field which captures active phase separation~\cite{Wittkowski2014, Nardini2017}, and (ii)~the coupled dynamics of scalar and vector fields which describes the emergence of motile deformable droplets~\cite{Tjhung2012, Tjhung2015}. Spatial decomposition has revealed that there is reduced heat rate at interfaces, and we have analyzed the leading order of heat rate at weak noise in relation with emerging patterns. For motile droplets, we have also shown that the heat rate undergoes a discontinuous transition when the droplets either split or merge.

Our work provides the relevant framework to study how heat rate relates to emerging patterns at hydrodynamic scale. The map of heat rate indicates which parts of the system mainly dissipate energy to sustain nonequilibrium fluctuations. At variance with the map deduced from explicit EPR, which was studied in~\cite{Nardini2017}, the heat-rate profile allows one to decipher where the underlying degrees of freedom contribute to shape the emergent profile of active fields. Note that our framework relies on the assumption of linear deviation from equilibrium thermodynamics~\cite{Mazur}, which does not hold for all active systems. It would be interesting to consider chemical reactions beyond the linear assumption, for which a thermodynamically consistent framework has been proposed recently~\cite{Esposito2016, Rao2018}.

Among the active field theories encompassed by our framework, many of them describe living systems, for instance dense assemblies of cells~\cite{Czajkowski2018, Yeomans2019} and swarms of bacteria~\cite{Yeomans2012, Goldstein2013}. While previous experimental works have already evaluated the dissipation of either isolated molecular motors~\cite{Toyabe2010, Mizuno2018}, cilia and flagella~\cite{Battle2016}, tracers in living cells~\cite{Ahmed2016, Ahmed2018}, or {\it in vitro} cytoskeletal network~\cite{Murrell2018}, only little is known regarding where energy is dissipated in spatially extended living systems. Our work opens the door to establishing maps of heat rate in models of living matter, with a potential to relating high/low dissipation locations with specific biological functions. Moreover, our predictions for the overall heat rate could potentially be tested against experimental measurements of the energy dissipated by living systems, such as metabolic rates~\cite{Gillooly2001, Makarieva2008}, using for instance calorimetric techniques~\cite{Howard2019}.

From a broad perspective, our framework lays the groundwork to bridge the gap between the thermodynamics of microscopic and hydrodynamic active theories. The stochastic thermodynamics of particle-based active dynamics has already received much attention in the last few years~\cite{Speck2016, Nardini2016, Mandal2017, Maggi2017, Shankar2018, Seifert2018, Bo2019}. Using systematic coarse-graining procedures, some active field theories are derived from microscopic equations, yielding explicit correspondences between hydrodynamic kinetic coefficients and microscopic parameters~\cite{Bertin2009, Farrell2012, Speck2013, Tailleur2013}. Based on these theories, our work offers the opportunity to compare the predictions for the heat rate of particle-based dynamics and that of their hydrodynamic counterparts, as way to analyze critically the energetics of active models at different scales.
Interestingly, a specific class of active models has considered explicitly the coupling between particle degrees of freedom and underlying chemical reactions, following the recipe of LIT~\cite{Kapral2018, Gaspard2019}. It would be interesting to explore whether coarse-graining this microscopic model leads to our framework at the hydrodynamic level. 

Moreover, one could examine the performances of work extraction for continuum models~\cite{Yeomans2016,Vitelli2019} and compare them with results obtained recently for particle-based engines~\cite{Martin2018, Pietzonka2019, Ekeh2020}. Furthermore, changing heat rate by using dynamical bias, one could study dynamical phase transitions in hydrodynamic theories, and compare them with that reported in active particles~\cite{Suma2017, Nemoto2018, Suri2019, Suri2020, GrandPre2020}. One expects the collective states of particle-based models emerging at high/low heat rate to coincide, at least qualitatively, with instabilities of hydrodynamic models in the same regime. If not, one could potentially try and revise the hydrodynamic equations to find a better agreement with their microscopic counterparts. Following this route, our work not only opens the door to controlling heat rate in active field theories, it also potentially provides a way to constrain their formulation. This calls for deeper investigations and encourages further contributions to the thermodynamics of active matter.


\acknowledgements{The authors acknowledge insightful discussions with Yongjoo Baek and Robert L. Jack. Work funded in part by the European Research Council under the EU's Horizon 2020 Programme, grant number 740269, and by the National Science Foundation under Grant No. NSF PHY-1748958. TM acknowledges support from the Blavatnik Postdoctoral Fellowship programme and the National Science Foundation Center for Theoretical Biological Physics (Grant PHY-2019745). \'EF acknowledges support from an ATTRACT Investigator Grant of the Luxembourg National Research Fund, an Oppenheimer Research Fellowship from the University of Cambridge, and a Junior Research Fellowship from St Catharine's College. MEC is funded by the Royal Society.}


\appendix

\section{Nonequilibrium grand-canonical ensemble}\label{app:grand}

Consider a simple model of a conserved field dynamics as presented in Sec.~\ref{sec:dyn} and depicted in Fig.~\ref{fig1}. This system can be described by the dynamics of three species: active particles ($\phi$) that are only present in the active subsystem, fuel ($n_f$), and products of the fuel consumption by the active particles ($n_p$). 
The three corresponding continuity equations are:
\begin{eqnarray}
\label{eq:grand1}
\nonumber& &\dot\phi + {\nabla} \cdot{\bf J} = 0 \, , \\
\nonumber& &\dot n_f + {\nabla} \cdot{\bf J}_f = -r \, , \\
& &\dot n_p + {\nabla} \cdot{\bf J}_p = r \, ,
\end{eqnarray}
where ${\bf J}_{f,p} = - \underline{\underline{\bf D}}^{\{f,p\}} \cdot \nabla \mu_{\{f,p\}}$,
and $r$ is the rate of fuel consumption that is non-vanishing only within the active subsystem. Here, the chemical potentials of the fuel and products are defined as usual $\mu_{\{f,p\}} = \delta {\cal F} / \delta n_{\{f,p\}}$. We continue by defining the chemical coordinates $n = \left(n_p - n_f\right)/2$ and $n_t = \left(n_p + n_f\right)/2$ such that $\mu_{f,p} = \left(\delta{\cal F} / \delta n_t \mp \delta{\cal F} / \delta n \right) / 2$, and the chemical potential difference is $\Delta\mu = \mu_f - \mu_p = - \delta{\cal F} / \delta n$. 
When diffusion of fuel/products is fast enough compared to the rate of fuel consumption $r$ and the dynamics of the active fields $\phi$, $\mu_{\{f,p\}}$ adjusts very fast compared with the active dynamics, so that it can be considered to be constant throughout the entire system. In such a case $\dot{n}_p = -\dot{n}_f = r$, $\dot{n}_t=0$, and 
\begin{eqnarray}
\label{eq:grand3}
\dot{n} = r\, ,
\end{eqnarray}
within the active subsystem. Although the dynamics of the chemical coordinate $n$ and the fuel/products are essentially the same, these fields are not equivalent. Specifically, the free-energy dependence on either fuel, products, or the chemical coordinates is generally different.

Finally, we assume that the timescale in which a significant change in $\Delta\mu$ occurs is very long compared to the timescales of interest. Then, the reservoirs of fuel/products can be regarded as having constant chemical potentials, and a constant chemical potential difference $\Delta\mu$ is maintained throughout the small active subsystem (see also Fig.~\ref{fig1}), which is the source of activity. This is what used in the main text Eq.~\eqref{eq:dyn_chem_g}. 
The construction described above essentially forms a {\it non-equilibrium grand-canonical ensemble}. Within this ensemble, $\dot{n}$ must be thought of as being the rate of fuel consumption, while the connection to the free energy of the reservoirs is seemingly lost.

\begin{figure}
	\centering
	\includegraphics[width=1\linewidth]{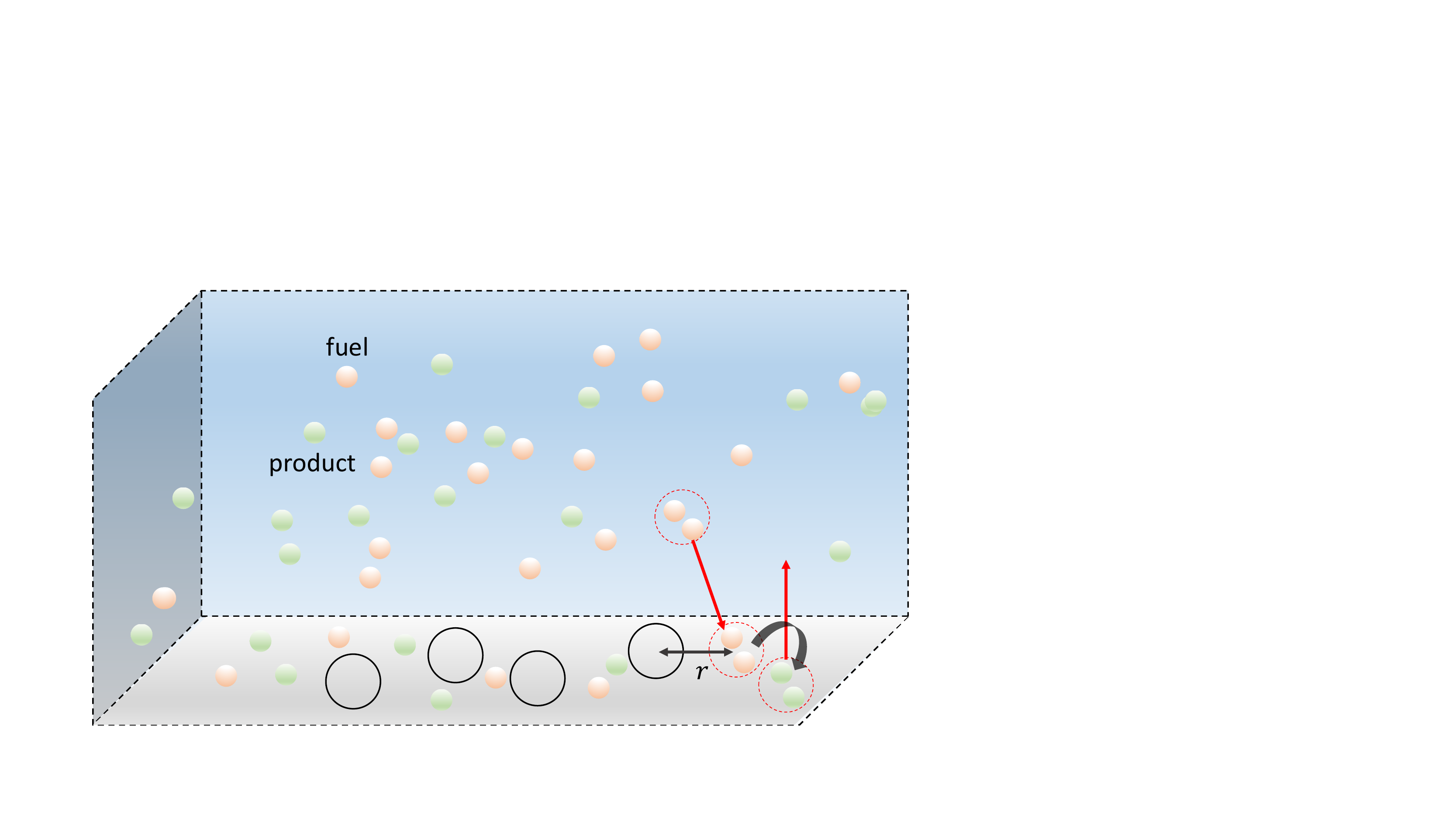}
	\caption{\label{fig7}
		Schematic of the active slab system. Small orange/green spheres are fuel/products and large hollow spheres are the active particles, which are confined to the surface and consume fuel at rate $r$. The products of the reaction diffuse out of the slab to the bulk system (reservoir of both fuel and products) while fuel molecules diffuse into the slab from the bulk to maintain constant chemical potential. 
	}
\end{figure}

A concrete example for such an active subsystem, which is also a prototype, is a thin slab as depicted in Fig.~\ref{fig7}. For instance, this would be an appropriate description for the experiment on light-activated self-propelled colloids of Ref.~\cite{Palacci2013}. In this geometry one may write $\underline{\underline{\bf D}}^{\{f,p\}}  = D_\perp^{\{f,p\}} \hat{\bf e}_\perp \hat{\bf e}_\perp+ D_\parallel^{\{f,p\}} \left(\underline{\underline{\bf I}} - \hat{\bf e}_\perp \hat{\bf e}_\perp\right)$ where $\hat{\bf e}_\perp$ refers to the direction perpendicular to the thin slab. Because the slab is thin, diffusion of particles in/out of the slab is much faster than within it, such that fuel/products do not flow within the slab, ${\bf J}_{f,p} \simeq \left({\bf J}_{f,p} \cdot  \hat{\bf e}_\perp\right)\hat{\bf e}_\perp$. Conservation of mass then dictates that ${\bf J}_p \cdot \hat{\bf e}_\perp = - {\bf J}_f \cdot \hat{\bf e}_\perp$ (the active particles cannot leave the slab) so that
\begin{eqnarray}
\label{eq:grand2}
\dot{n}_p = -\dot{n}_f =  - \nabla_{\hat{\bf e}_\perp} \left( {\bf J}_p \cdot \hat{\bf e}_\perp\right) +  r \, .
\end{eqnarray}
When diffusion of fuel/products in/out of the slab is fast compared with the active dynamics we get Eq.~\eqref{eq:grand3} within the slab. Note that in this example the diffusion of fuel/products within the slab does not need to be fast compared with the active dynamics. It is sufficient to have fast diffusion of fuel/products perpendicular to the slab.

At times long enough that the fuel reservoir starts to become exhausted, one must consider the change of $\Delta\mu$ due to fuel/products fluxes in/out of the active subsystem, as in Eq.~\eqref{eq:grand2}. On these timescales  there should not be any steady-state heat production. This is evident from Eq.~\eqref{eq:entropy_b} after substituting $\Delta\mu = -\delta{\cal F} / \delta n$, $\dot{n}_t=0$ and Eq.~\eqref{eq:grand3}, which gives $\dot{\cal Q} = {\rm d} \langle {\cal F}[n,n_t]\rangle / {\rm d} t = 0$.


\section{Spurious drift terms}\label{app:discretisation}

In this Appendix, we obtain the expression of the spurious drift terms $\{{\boldsymbol\nu_\Omega},\chi_\Omega\}$ in~(\ref{eq:dyn_phi}-\ref{eq:dyn_chem}) and~(\ref{eq:dyn_field}-\ref{eq:dyn_chem_b_g}). To this aim, we first derive the Fokker-Planck equations (FPEs) associated with the spatially-discretized dynamics. Then, we choose the spurious drift terms so that the Boltzmann distribution is the steady state solution of FPEs in the equilibrium regime. We focus on the one-dimensional case for simplicity ($d=1$), since the generalization to higher $d$ is straightforward. To generalize the discussion to $d$ dimensions, one only needs to use the $d$-dimensional version of the gradient matrix instead of the one-dimensional matrix used below.

The spatial discretization amounts to considering the variables $\{\phi_i(t), p_i(t)\}$, where the indices $i$ denote lattice coordinates, whose dynamics converge to that of $\{\phi(x,t), p(x,t)\}$ in the limit of small lattice constant $\Delta x$, where $x=i\Delta x$. In particular, we introduce the gradient matrix $\bf A$ defined by
\begin{equation} \label{eq:derivative}
	\begin{aligned}
		\lim_{\Delta x\to0}\sum_j A_{ij}\phi_j(t) &= \partial_x \phi(x,t) ,
		\\
		\lim_{\Delta x\to0} \,\frac{1}{\Delta x}\sum_k A_{ik} \frac{\partial}{\partial\phi_k(t)} &= \partial_x \frac{\delta}{\delta\phi(x,t)} .
	\end{aligned}
\end{equation}
A standard choice for $\bf A$ is given by $A_{ij} = (\delta_{i,j-1}-\delta_{i,j+1})/(2\Delta x)$, though other spatial discretizations are possible. In what follows, we discuss the consequence of such a choice in the form of the spurious drift terms.

\subsection{Conserved dynamics for scalar field}\label{app:spur}

We first consider the conserved dynamics for a scalar field in~(\ref{eq:dyn_phi}-\ref{eq:dyn_chem}), and write them in the discretized form as
\begin{equation}\label{eq:dyn_disc}
\begin{aligned}
\dot \phi_i &= \sum_j A_{ij} \Big( \lambda\sum_k A_{jk}\psi_k - \Delta\mu_j C_j  - T \nu_j - \Lambda_j \Big) ,
\\
\dot n_i &= \gamma\Delta\mu_i - C_i \sum_j A_{ij}\psi_j + T \chi_i + \xi_i ,
\end{aligned}
\end{equation}
where $\psi_i = (\delta{\cal F}/\delta\phi) (x=i\Delta x)$, and the coupling term $C_i = C(\phi_i, \sum_j A_{ij}\phi_j, \dots)$ depends on $\phi$ and its gradients in general. The noise terms $\{\Lambda_i,\xi_i\}$ are Gaussian with zero mean and correlations given by
\begin{equation}\label{eq:corr}
\big\langle \big[\Lambda_i,\xi_i\big](t) \big[\Lambda_j,\xi_j\big]^\intercal(0) \big\rangle = 2 T \,{\mathbb L}_i \frac{\delta_{ij}\delta(t)}{\Delta x} ,
\quad
{\mathbb L}_i =
\begin{bmatrix}
\lambda & C_i
\\
C_i & \gamma
\end{bmatrix}
.
\end{equation}
Given that the correlations between $\Lambda_i$ and $\xi_i$ depend on the variable $\phi_i$ through the coupling term $C_i$, one has to specify the temporal discretization scheme of~\eqref{eq:dyn_disc}. In what follows, and in the main text, we choose Stratonovich convention, which allows one to use the standard rules of differential calculus~\cite{Gardiner}. This is particularly convenient when deriving the expression of the heat rate $\dot{\cal Q}$ defined in~\eqref{eq:entropy}.

The associated FPE for the probability density $P(\{\phi_i,n_i\},t)$ can then be derived following standard methods as~\cite{Gardiner}
\begin{widetext}
\begin{equation}\label{eq:fpe}
	\begin{aligned}
		\dot P &= \sum_{i,j} A_{ij} \frac{\partial}{\partial\phi_i} \Big[\Big( - \lambda \sum_k A_{jk}\psi_k + \Delta\mu_j C_j + T\nu_j \Big) \,P \,\Big] + \sum_i \frac{\partial}{\partial n_i} \Big[\Big( - \gamma \Delta\mu_i + C_i \sum_j A_{ij}\psi_j - T \chi_i \Big) \,P\,\Big]
		\\
		&\quad + \frac{T}{\Delta x} \sum_{i,a,b,c} \bigg[ \sum_j A_{ij} \frac{\partial}{\partial\phi_j}, \frac{\partial}{\partial n_i}\bigg]_a {\mathbb M}_{i,ab} \bigg[ \sum_k A_{ik} \frac{\partial}{\partial\phi_k}, \frac{\partial}{\partial n_i}\bigg]^\intercal_c \big( {\mathbb M}_{i,cb} P \big) ,
	\end{aligned}
\end{equation}
\end{widetext}
where we have introduced the matrix ${\mathbb M}_i$ defined by ${\mathbb M}_i {\mathbb M}_i^\intercal = {\mathbb L}_i$. In the continuum limit of small $\Delta x$, it follows using~\eqref{eq:derivative} that~\eqref{eq:fpe} converges to the standard functional FPE for the probability density $P([\phi(x),n(x)],t)$~\cite{Tailleur2008, Solon2015}. Importantly, by taking $\{\nu_i,\chi_i\}$ as
\begin{equation}\label{eq:spurious}
\big[\nu_i, \chi_i\big]_a = \frac{1}{\Delta x} \sum_{b,c} {\mathbb M}_{i,ab} \bigg[ \sum_k A_{ik} \frac{\partial}{\partial\phi_k}, \frac{\partial}{\partial n_i}\bigg]_c {\mathbb M}_{i,cb} ,
\end{equation}
the stationary solution of~\eqref{eq:fpe} is given by the Boltzmann distribution $P_{\rm s}\sim{\rm e}^{- \Delta x \,F/T}$ at equilibrium, namely when $[\psi_i, \Delta\mu_i] = [ \partial F/\partial\phi_i, -\partial F/\partial n_i]$, as expected~\cite{Chaikin, Lau2007}. As a result, the expression of $\{\nu_i,\chi_i,{\mathbb L}_i\}$ in~\eqref{eq:corr} and~\eqref{eq:spurious} provide a systematic way to compute the spurious drift terms in terms of $C_i$. When $C_i$ is independent of $n_i$, as is assumed below,~\eqref{eq:spurious} vanishes if $C_i$ only depends on $\phi_i$, namely when it is a local function of $\phi$ independent of its gradients. Besides, the extension of~\eqref{eq:spurious} for $d>1$ follows directly by substituting the $d$-dimensional version of the gradient matrix $\bf A$.

When $d=1$, the chain rule
\begin{equation}
\frac{\partial {\mathbb M}_{i,ab}}{\partial\phi_j} = \frac{\partial {\mathbb M}_{i,ab}}{\partial C_i} \,\frac{\partial C_i}{\partial\phi_j} ,
\end{equation}
then leads to simplify~\eqref{eq:spurious} as
\begin{equation}\label{eq:spurious_g}
\begin{aligned}
\nu_i &= \frac{1}{\Delta x} \bigg( {\mathbb M}_{i,11} \frac{\partial {\mathbb M}_{i,11}}{\partial C_i} + {\mathbb M}_{i,12} \frac{\partial {\mathbb M}_{i,12}}{\partial C_i} \bigg) \sum_j A_{ij} \frac{\partial C_i}{\partial\phi_j} ,
\\
\chi_i &= \frac{1}{\Delta x} \bigg( {\mathbb M}_{i,21} \frac{\partial {\mathbb M}_{i,11}}{\partial C_i} + {\mathbb M}_{i,22} \frac{\partial {\mathbb M}_{i,12}}{\partial C_i} \bigg) \sum_j A_{ij} \frac{\partial C_i}{\partial\phi_j} .
\end{aligned}
\end{equation}
The matrix ${\mathbb M}_i$ can be written as ${\mathbb M}_i = {\mathbb P}_i^{-1}{\mathbb D}_i{\mathbb P}_i$, where
\begin{equation}
\begin{aligned}
{\mathbb D}_i &=
\begin{bmatrix}
\sqrt{\tau_{i,-}} & 0
\\
0 & \sqrt{\tau_{i,+}}
\end{bmatrix}
,
\quad
{\mathbb P}_i =
\begin{bmatrix}
(\tau_{i,-} - \gamma)/C_i & 1
\\
(\tau_{i,+} - \gamma)/C_i & 1
\end{bmatrix}
,
\\
\tau_{i,\pm} &= \frac{1}{2} \bigg[ \gamma+\lambda \pm \sqrt{4 C_i^2 + (\gamma-\lambda)^2} \bigg] .
\end{aligned}
\end{equation}
Substituting the expression of ${\mathbb M}_i$ in~\eqref{eq:spurious_g}, we deduce that $\nu_i$ always vanishes for any $C_i$ in $d=1$, yet it can still potentially be non-zero in higher dimensions. Besides, we deduce the expression of $\chi_i$ as
\begin{equation}\label{eq:pi}
\chi_i = \frac{1}{\Delta x} \,\frac{2C_i^2+(\gamma-\lambda)\Big[\gamma-\sqrt{\gamma\lambda-C_i^2}\Big]}{4C_i^2+(\gamma-\lambda)^2} \sum_j A_{ij} \frac{\partial C_i}{\partial\phi_j} .
\end{equation}
To obtain Eq.~\eqref{eq:entropy_c_bis} from Eq.~\eqref{eq:entropy_c}, one has to evaluate $\sum_i\big\langle C_i\Lambda_i\big\rangle$ following standard stochastic calculus~\cite{Gardiner}, which reads
\begin{equation}\label{eq:C_lam}
	\begin{aligned}
		\sum_i\big\langle C_i\Lambda_i\big\rangle	= T \sum_{i,j} &A_{ij} \,\bigg\langle {\mathbb M}_{i,11} \frac{\partial}{\partial\phi_j} \big({\mathbb M}_{i,11} C_i\big)
		\\
		&\quad\, + {\mathbb M}_{i,12} \frac{\partial}{\partial\phi_j} \big({\mathbb M}_{i,12} C_i\big) \bigg\rangle ,
	\end{aligned}
\end{equation}
where we have used again that $C_i$ is independent of $n_i$. From~(\ref{eq:pi}-\ref{eq:C_lam}), it follows that the relation between the heat rate $\dot{\cal Q}$ and the explicit entropy production rate $\cal S$ given in~\eqref{eq:entropy_c_bis} holds whenever $\sum_j A_{ij} (\partial C_i/\partial\phi_j)=0$, which is the case considered in the main text.

Let us focus on the specific coupling term $C_{\rm AMB} = \partial_x(\partial_x\phi)^2 = 2 (\partial_x\phi)\partial_{xx}^2\phi$ corresponding to Active Model B, as considered in Sec.~\ref{sec:phase}. This coupling term can be written using different discretization schemes, such as
\begin{equation}\label{eq:Cphi-discretized}
\begin{aligned}
C_i^{(1)} &= \sum_{k,l,m} A_{ik} (A_{kl}\phi_l)(A_{km}\phi_m) ,
\\
C_i^{(2)} &= 2 \sum_{k,l,m} (A_{ik} \phi_k) (A_{il} A_{lm}\phi_m) ,
\end{aligned}
\end{equation}
both of which converge to $C_{\rm AMB}$ at small $\Delta x$. A priori, one might expect the spurious drift terms to be independent of the discretization scheme, yet we now show that different discretizations yield different expressions for the spurious drift terms in general. For $C^{(1)}$, we get
\begin{equation}\label{eq:spur_1}
\begin{aligned}
\sum_j A_{ij} \frac{\partial C_i^{(1)}}{\partial\phi_j} &= 2 \sum_{j,k,l} A_{ij} A_{ik} A_{kj} A_{kl} \phi_l
\\
&= - 2 \sum_{j,k,l} (A_{ij} A_{jk}) A_{ik} A_{kl} \phi_l
\\
&= - 2 \sum_{k,l} \big[{\bf A}^2\big]_{ik} A_{ik} A_{kl} \phi_l ,
\end{aligned}
\end{equation}
where we have used $A_{ij}=-A_{ji}$. Taking $A_{ij} = (\delta_{i,j-1} - \delta_{i,j+1})/(2\Delta x)$, we deduce $[{\bf A}^2]_{ik} A_{ik}=0$, so that~\eqref{eq:spur_1} is zero. Substituting~\eqref{eq:spur_1} in~\eqref{eq:spurious_g}, we conclude that there is no spurious drift associated with $C^{(1)}$ since both $\nu_i$ and $\chi_i$ vanish, yet this no longer holds when considering higher-order schemes for the gradient matrix $\bf A$. For $C^{(2)}$, we get
\begin{equation}\label{eq:spur_2}
\begin{aligned}
\sum_j &A_{ij} \frac{\partial C_i^{(2)}}{\partial\phi_j}
\\
&= 2 \sum_{j,k,l} A_{ij} \big( A_{ij} A_{il} A_{lk} + A_{il} A_{lj} A_{ik} \big) \phi_k
\\
&= - 2 \sum_{j,k,l} \big( (A_{ij}A_{ji}) (A_{il}A_{lk}) + (A_{il}A_{lj}A_{ji}) A_{ik} \big) \phi_k
\\
&= - 2 \sum_k \big( \big[{\bf A}^2\big]_{ii} \big[{\bf A}^2\big]_{ik} + \big[{\bf A}^3\big]_{ii} A_{ik} \big) \phi_k ,
\end{aligned}
\end{equation}
where we used again $A_{ij}=-A_{ji}$. Given that ${\bf A}$ is anti-symmetric, any odd (even) power of ${\bf A}$ is anti-symmetric (symmetric), so that $[{\bf A}^3]_{ii}=0$ and $[{\bf A}^2]_{ii}\neq 0$. Then,~\eqref{eq:spur_2} is always non-zero for any form of the gradient matrix $\bf A$. The examples in~(\ref{eq:spur_1}-\ref{eq:spur_2}) illustrate that the choice of spatial discretization affects drastically the form of the spurious drift terms.

For the study of Active Model B presented in Sec.~\ref{sec:phase}, we choose to discretize $C_{\rm AMB}$ using $C^{(1)}$. Since the corresponding spurious drift terms vanish, this choice allows us to embed Active Model B in a thermodynamically consistent framework without need to change the dynamical equations considered in~\cite{Wittkowski2014, Nardini2016}.


\subsection{Generalized field dynamics}\label{app:spur_gen}

We now consider the generalized dynamics for conserved and non-conserved fields in~(\ref{eq:dyn_field}-\ref{eq:dyn_chem_b_g}). When the driving parameter is a chemical potential difference ($\Delta_\Omega = \Delta\mu_\Omega$), the spurious drift terms follow from a straightforward extension of~\eqref{eq:spurious} as
\begin{equation}\label{eq:spurious_b}
	\begin{aligned}
		\big[\nu_{\phi,i}, \chi_{\phi,i}\big]_a &= \frac{1}{\Delta x} \sum_{b,c} {\mathbb M}^{(\phi)}_{i,ab} \bigg[ \sum_k A_{ik} \frac{\partial}{\partial\phi_k}, \frac{\partial}{\partial n_{\phi,i}}\bigg]_c {\mathbb M}^{(\phi)}_{i,cb} ,
		\\
		\big[\nu_{p,i}, \chi_{p,i}\big]_a &= \frac{1}{\Delta x} \sum_{b,c} {\mathbb M}^{(p)}_{i,ab} \bigg[ \frac{\partial}{\partial p_i}, \frac{\partial}{\partial n_{p,i}}\bigg]_c {\mathbb M}^{(p)}_{i,cb} ,
	\end{aligned}
\end{equation}
where 
\begin{equation}\label{eq:corr_b}
	{\mathbb M}^{(\Omega)}_i \Big[{\mathbb M}^{(\Omega)}_i\Big]^\intercal = {\mathbb L}^{(\Omega)}_i ,
	\quad
	{\mathbb L}^{(\Omega)}_i =
	\begin{bmatrix}
		\lambda_\Omega & C_{\Omega,i}
		\\
		C_{\Omega,i} & \gamma_\Omega
	\end{bmatrix}
	.
\end{equation}
The expression of $\{\nu_\Omega,\chi_\Omega,{\mathbb L}_\Omega\}$ in~(\ref{eq:spurious_b}-\ref{eq:corr_b}) can then be used to derive explicitly the spurious drift terms for given coupling terms ${\bf C}_\Omega$. As discussed in Sec.~\ref{app:spur}, the choice for spatial discretization of the gradient terms appearing in ${\bf C}_\Omega$ is crucial to determine the corresponding spurious drift terms: A judicious choice can potentially make $\{\nu_\Omega,\chi_\Omega\}$ vanish.

The case where the driving parameter represents a chemical current ($\Delta_\Omega = \dot n_\Omega/\gamma_\Omega$) deserves a more careful treatment, which we discuss now. For simplicity, we address the dynamics of a polar field $p$ without any conserved scalar field $\phi$, as given by
\begin{equation}\label{eq:dyn_disc_b}
\begin{aligned}
\dot p_i &= \lambda h_i + (\dot n_i/\gamma) \, C_i + T \nu_i + \Lambda_i ,
\\
\dot n_i &= \gamma\Delta\mu_i + C_i\, h_i + T \chi_i + \xi_i ,
\end{aligned}
\end{equation}
where $h_i = - (\delta{\cal F}/\delta p)(x=i\Delta x)$, and $\{\Lambda_i,\xi_i\}$ are Gaussian noises with zero mean and correlations
\begin{equation}
\big\langle \big[\Lambda_i,\xi_i\big](t) \big[\Lambda_j,\xi_j\big]^\intercal(0) \big\rangle = 2 T
\begin{bmatrix}
\lambda & 0
\\
0 & \gamma
\end{bmatrix}
\frac{\delta_{ij}\delta(t)}{\Delta x} .
\end{equation}
Note that there is no longer any correlation between $\Lambda_i$ and $\xi_i$ in contrast with~\eqref{eq:corr}. To derive the corresponding FPE for $P(\{p_i, n_i\},t)$, it is convenient to substitute the expression of $\dot n_i$ in the dynamics of $p_i$, yielding
\begin{equation}
\dot p_i = \bigg(\lambda + \frac{C_i^2}{\gamma}\bigg) h_i + \Delta\mu_i\, C_i + T \bigg(\nu_i + \frac{C_i\chi_i}{\gamma}\bigg) + \Gamma_i .
\end{equation}
Here the noise term $\Gamma_i = \Lambda_i + C_i\xi_i/\gamma$ is Gaussian with zero mean and correlations given by
\begin{equation}\label{eq:corr_c}
\big\langle \big[\Gamma_i,\xi_i\big](t) \big[\Gamma_j,\xi_j\big]^\intercal(0) \big\rangle = 2 T \,{\mathbb K}_i \frac{\delta_{ij}\delta(t)}{\Delta x} ,
\end{equation}
where the Onsager matrix ${\mathbb K}_i$ reads
\begin{equation}\label{eq:corr_c1}
\begin{aligned}
{\mathbb K}_i =
\begin{bmatrix}
\lambda + C_i^2/\gamma & C_i
\\
C_i & \gamma
\end{bmatrix}
.
\end{aligned}
\end{equation}
Note that the dynamics can be written in a compact form, analogous to that given in~\eqref{eq:onsager} when the driving parameter is a chemical potential difference, as
\begin{equation}\label{eq:dyn_disc_c}
\big[\dot p_i, \dot n_i \big] = {\mathbb K}_i \big[h_i, \Delta\mu_i \big] + T \big[\nu_i+C_i\chi_i/\gamma, \chi_i\big] + \big[\Gamma_i, \xi_i\big] .
\end{equation}
Though $\dot n$ is taken constant in~\eqref{eq:dyn_disc_c}, we consider the case where both $p_i$ and $n_i$ are stochastic variables to obtain the expression of the spurious drift terms $\{\nu_i,\chi_i\}$. The FPE for $P(\{p_i, n_i\},t)$ then reads
\begin{widetext}
\begin{equation}\label{eq:fpe_b}
	\begin{aligned}
		\dot P &= \sum_i \frac{\partial}{\partial p_i} \bigg\{\bigg[ - \bigg(\lambda + \frac{C_i^2}{\gamma} \bigg) h_i - \Delta\mu_i C_i - T \bigg(\nu_i+\frac{C_i\chi_i}{\gamma}\bigg) \bigg] \,P \,\bigg\} + \sum_i \frac{\partial}{\partial n_i} \Big[\Big( - \gamma \Delta\mu_i - C_i h_i - T \chi_i \Big) \,P\,\Big]
		\\
		&\quad + \frac{T}{\Delta x} \sum_{i,a,b,c} \bigg[ \frac{\partial}{\partial p_i}, \frac{\partial}{\partial n_i}\bigg]_a {\mathbb J}_{i,ab} \bigg[ \frac{\partial}{\partial p_i}, \frac{\partial}{\partial n_i}\bigg]^\intercal_c \big( {\mathbb J}_{i,cb} P \big) ,
	\end{aligned}
\end{equation}
\end{widetext}
where we have introduced the matrix ${\mathbb J}_i$ defined by ${\mathbb J}_i {\mathbb J}^\intercal_i={\mathbb K}_i$. Choosing the spurious drift terms
\begin{equation}\label{eq:spurious_c}
\bigg[\nu_i+\frac{C_i\chi_i}{\gamma}, \chi_i\bigg]_a = \frac{1}{\Delta x} \sum_{b,c} {\mathbb J}_{i,ab} \bigg[\frac{\partial}{\partial p_i}, \frac{\partial}{\partial n_i}\bigg]_c {\mathbb J}_{i,cb} 
\end{equation}
enforces that the stationary solution of~\eqref{eq:fpe_b} is $P_{\rm s}\sim{\rm e}^{- \Delta x \,F/T}$ when $[h_i, \Delta\mu_i] = [ -\partial F/\partial h_i, -\partial F/\partial n_i] $, as expected in equilibrium~\cite{Chaikin, Lau2007}. Then,~(\ref{eq:corr_c}-\ref{eq:spurious_c}) define $\{\nu_i,\chi_i\}$ in terms of $C_i$ through ${\mathbb J}_i$ when the driving parameter represents a chemical current ($\Delta_p=\dot n/\gamma$). This definition is in general different from that given in~(\ref{eq:spurious_b}-\ref{eq:corr_b}) when the driving parameter is a chemical potential ($\Delta_p=\Delta\mu_p$).

For $d=1$, the spurious drift terms $\{\nu_i,\chi_i\}$ are proportional to $\partial C_i/\partial p_i$. In particular, taking
\begin{equation}\label{eq:Cp-discretized}
C^{(1)}_{p,i} = p_i \sum_j A_{ij} p_j ,
\quad
C^{(2)}_{p,i} = \frac{1}{2} \sum_j A_{ij} p_j^2 ,
\end{equation}
which both converge to $C_p=p\partial_x p = (1/2) \partial_x p^2$ at small $\Delta x$, as considered in Sec.~\ref{sec:drop}, we get
\begin{equation}
\frac{\partial C^{(1)}_{p,i}}{\partial p_i} = \sum_j A_{ij} p_j ,
\quad
\frac{\partial C^{(2)}_{p,i}}{\partial p_i} = 0 ,
\end{equation}
where we have used $A_{ii}=0$. It follows that $\{\nu_i,\chi_i\}$ vanish for the coupling term $C^{(2)}_p$ but not for $C^{(1)}_p$. As already noticed for the  conserved dynamics of a scalar field, see~(\ref{eq:spur_1}-\ref{eq:spur_2}), different spatial discretizations yield different spurious drift terms. For the motile polar droplets studied in Sec.~\ref{sec:drop}, we take the coupling term to be discretized as $C_p^{(2)}$ which leads to vanishing spurious drift terms.  Note that
\begin{equation}
	\begin{aligned}
		\sum_i \big\langle C_{p,i} \Lambda_{p,i} \big\rangle = T &\sum_i \bigg\langle {\mathbb J}_{i,11} \frac{\partial}{\partial p_i} \big({\mathbb J}_{i,11} C_{p,i}\big)
		\\
		&\quad\; + {\mathbb J}_{i,12} \frac{\partial}{\partial p_i} \big({\mathbb J}_{i,12} C_{p,i}\big) \bigg\rangle
	\end{aligned}
\end{equation}
also vanishes whenever $\partial C_{p,i} / \partial p_i = 0$, in particular it does so for $C_p^{(2)}$.

Moreover, in the case where the dynamics~\eqref{eq:dyn_disc_b} features an additional conserved scalar field with driving parameter proportional to chemical current, namely $\Delta_\phi = \dot n_\phi/\gamma_\phi$, the spurious drift terms follow straightforwardly by extending~\eqref{eq:spurious_c}. Indeed, since the FPE for $P(\{\phi,p_i,n_{\phi,i},n_{p,i}\},t)$ can be separated into two sectors, associated with derivatives given by either $\{\partial/\partial p_i, \partial/\partial n_{p,i}\}$ or $\{\sum_j A_{ij}(\partial/\partial\phi_j), \partial/\partial n_{\phi,i}\}$, we get
\begin{equation}
\begin{aligned}
&\bigg[\nu_{\phi,i}+\frac{C_{\phi,i}\chi_{\phi,i}}{\gamma_\phi}, \chi_{\phi,i}\bigg]_a
\\
&\quad = \frac{1}{\Delta x} \sum_{b,c} {\mathbb J}^{(\phi)}_{i,ab} \bigg[\sum_k A_{ik}\frac{\partial}{\partial\phi_k}, \frac{\partial}{\partial n_i}\bigg]_b {\mathbb J}^{(\phi)}_{i,cb} ,
\end{aligned}
\end{equation}
where
\begin{equation}
{\mathbb J}^{(\phi)}_i \Big[{\mathbb J}^{(\phi)}_i\Big]^\intercal = {\mathbb K}^{(\phi)}_i ,
\quad
{\mathbb K}^{(\phi)}_i =
\begin{bmatrix}
\lambda_\phi + C_{\phi,i}^2/\gamma_\phi & C_{\phi,i}
\\
C_{\phi,i} & \gamma_\phi
\end{bmatrix}
.
\end{equation}
The spurious drift terms vanish when $C_\phi$ depends on $\phi$ only locally. In particular, this is the case for $C_\phi = \phi p$ as considered in Sec.~\ref{sec:drop}.


\section{Heat rate}\label{app:entropy}

This Appendix is devoted to deriving the heat rate $\dot{\cal Q}$, as defined in~\eqref{eq:entropy}, for the dynamics~(\ref{eq:dyn_field}-\ref{eq:dyn_chem_b_g}). We obtain a generic expression in terms of the driving parameter and its conjugated chemical field. Moreover, we derive explicitly the dependence of $\dot{\cal Q}$ on model parameters for active phase separation and motile polar droplets, to leading order in noise strength, as considered respectively in Secs.~\ref{sec:phase} and~\ref{sec:drop}.

\subsection{Generalized field dynamics}

We first consider a generalized dynamics for a conserved scalar field $\phi$ and a polar field $\bf p$ of the form
\begin{equation}\label{eq:dyn_gen}
\begin{aligned}
\dot\phi &= -\nabla\cdot{\bf J} ,
\\
\begin{bmatrix}
{\bf J} \\ \;\dot n_\phi\; \\ \dot{\bf p} \\ \dot n_p
\end{bmatrix}
&= {\mathbb L}
\begin{bmatrix}
\;-\nabla(\delta\cal F/\delta\phi)\; \\ \Delta\mu_\phi \\ - \delta\cal F/\delta{\bf p} \\ \Delta\mu_p
\end{bmatrix}
+ T
\begin{bmatrix}
\;{\boldsymbol\nu}_\phi\; \\ \chi_\phi \\ {\boldsymbol\nu}_p \\ \chi_p
\end{bmatrix}
+
\begin{bmatrix}
\;{\boldsymbol\Lambda}_\phi\; \\ \xi_\phi \\ {\boldsymbol\Lambda}_p \\ \xi_p
\end{bmatrix}
,
\end{aligned}
\end{equation}
where the noise term ${\boldsymbol\Xi} = [{\boldsymbol\Lambda}_\phi, \xi_\phi, {\boldsymbol\Lambda}_p, \xi_p]$ is Gaussian with zero mean and correlations given by
\begin{equation}
\big\langle {\boldsymbol\Xi}({\bf r},t) {\boldsymbol\Xi}^\intercal({\bf r}',t') \big\rangle = 2 T \,{\mathbb L}({\bf r},t) \delta({\bf r}-{\bf r}') \delta(t-t') ,
\end{equation}
and the spurious drift terms are given for $d=1$ as
\begin{equation}\label{eq:dyn_gen_b}
\begin{aligned}
&\big[ \nu_{\phi,i}, \chi_{\phi,i}, \nu_{p,i}, \chi_{p,i} \big]_a
\\
&= \frac{1}{\Delta x} \sum_{b,c} {\mathbb M}_{i,ab} \bigg[ \sum_k A_{ik} \frac{\partial}{\partial\phi_k}, \frac{\partial}{\partial n_{\phi,i}}, \frac{\partial}{\partial p_i}, \frac{\partial}{\partial n_{p,i}}\bigg]_c {\mathbb M}_{i,cb} ,
\end{aligned}
\end{equation}
where $\mathbb M$ is defined by ${\mathbb M}{\mathbb M}^\intercal = {\mathbb L}$. In contrast with~(\ref{eq:dyn_field}-\ref{eq:dyn_chem_b_g}), we now consider an arbitrary Onsager matrix $\mathbb L$, with the only constraint that it should be positive semi-definite ($\det{\mathbb L}\geq0$).

Following~\cite{Lau2007, Lecomte2017}, the path probability ${\cal P}\sim{\rm e}^{-{\cal A}}$ associated with~(\ref{eq:dyn_gen}-\ref{eq:dyn_gen_b}) is defined by
\begin{equation}\label{eq:action_field}
\begin{aligned}
{\cal A} &= \frac{1}{4T} \int_0^t \int_V
\left(
\begin{bmatrix}
{\bf J} \\ \;\dot n_\phi\; \\ \dot{\bf p} \\ \dot n_p
\end{bmatrix}
+ {\mathbb L}
\begin{bmatrix}
\;\nabla(\delta\cal F/\delta\phi)\; \\ -\Delta\mu_\phi \\ \delta\cal F/\delta{\bf p} \\ -\Delta\mu_p
\end{bmatrix}
\right)
\\
&\quad \times
{\mathbb L}^{-1}
\left(
\begin{bmatrix}
{\bf J} \\ \;\dot n_\phi\; \\ \dot{\bf p} \\ \dot n_p
\end{bmatrix}
+ {\mathbb L}
\begin{bmatrix}
\;\nabla(\delta\cal F/\delta\phi)\; \\ -\Delta\mu_\phi \\ \delta\cal F/\delta{\bf p} \\ -\Delta\mu_p
\end{bmatrix}
\right)^\intercal
{\rm d}{\bf r}{\rm d} s \, ,
\end{aligned}
\end{equation}
where, as a consequence of the Stratonovich discretization, the spurious drift terms do not appear in the expression~\eqref{eq:action_field}~\cite{Lau2007}. 
Note that some terms which are even under time-reversal have not been written explicitly in~\eqref{eq:action_field} since they are not relevant for deriving the heat rate. These terms could potentially be relevant if one or several of the order parameters were odd under time reversal. The time-reversed dynamic action ${\cal A}^{\rm R}$ follows from~\eqref{eq:action_field} by changing the sign of $[{\bf J}, \dot n_\phi, \dot{\bf p}, \dot n_p]$. From the definition in~\eqref{eq:entropy}, the heat rate can be written as
\begin{equation}
\dot{\cal Q} = \lim_{t\to\infty}\frac{T}{t} \left<{\cal A}^{\rm R}-{\cal A}\right> ,
\end{equation}
yielding
\begin{equation}\label{eq:heat}
\begin{aligned}
\dot{\cal Q} &= \int_V \bigg\langle - {\bf J}\cdot\nabla\frac{\delta\cal F}{\delta\phi} - \dot{\bf p}\cdot\frac{\delta\cal F}{\delta\bf p} + \dot n_\phi\Delta\mu_\phi + \dot n_p\Delta\mu_p \bigg\rangle \,{\rm d}{\bf r}
\\
&= - \frac{{\rm d}\langle{\cal F}\rangle}{{\rm d}t} + \int_V \big\langle \dot n_\phi\Delta\mu_\phi + \dot n_p\Delta\mu_p \big\rangle \,{\rm d}{\bf r} ,
\end{aligned}
\end{equation}
where we have used $\dot\phi = -\nabla\cdot{\bf J}$. In steady state, using ${\rm d}\langle{\cal F}\rangle/{\rm d}t=0$, we then deduce the final expression of heat rate given in~\eqref{eq:entropy_d}.

To treat the case where the driving coefficients are odd ($\Delta_\Omega=\dot n_\Omega/\gamma_\Omega$), the first step is to substitute in the dynamics of $\{\phi,{\bf p}\}$ the corresponding expressions for $\dot n_\Omega$, as is done in Sec.~\ref{app:spur_gen}, for instance. Then, following the same procedure as in~(\ref{eq:action_field}-\ref{eq:heat}), it is straightforward to show that~\eqref{eq:entropy_d} also holds for odd driving coefficients.


\subsection{Active phase separation}

We proceed by considering the specific dynamics of active phase separation, as studied in Sec.~\ref{sec:phase}. The difference between $T\cal S$ and the non-trivial contribution to heat rate $\dot{\cal Q}-\gamma V\Delta\mu^2$, denoted by $\Delta$ in what follows, reads
\begin{equation}
	\Delta = 4 \Delta\mu^2 \int_V \big\langle \big(\partial_x \phi\big)^2 \big(\partial^2_x \phi\big)^2 \big\rangle \,{\rm d}x .
\end{equation}
Expanding the density field around the homogeneous profile $\phi_0(x)=\bar\phi$ as $\phi=\bar\phi+\sqrt{T}\phi_1+{\cal O}(T)$, we get
\begin{equation}
	\Delta = 4 (T\Delta\mu)^2 \int_V \big\langle \big(\partial_x \phi_1\big)^2 \big(\partial^2_x \phi_1\big)^2 \big\rangle \,{\rm d}x + {\cal O}(T^3) ,
\end{equation}
which, by introducing the Fourier coefficients $\tilde\phi_1(k) = (1/V) \int_V {\rm e}^{-{\rm i}kx} \phi_1(x) {\rm d}x$, can be written as
\begin{equation}\label{eq:delta}
	\begin{aligned}
		\Delta &= - 4 V (T\Delta\mu)^2 \sum_{k_1,k_2,k_3} k_1 k_2 k_3^2 (k_1+k_2+k_3)^2
		\\
		&\quad\times \big\langle \tilde\phi_1(k_1) \tilde\phi_1(k_2) \tilde\phi_1(k_3) \tilde\phi_1^*(k_1+k_2+k_3) \big\rangle + {\cal O}(T^3) ,
	\end{aligned}
\end{equation}
where $*$ denotes complex conjugate. Given that $\tilde\phi_1$ has Gaussian statistics with zero mean, Wick's theorem enforces that
\begin{equation}\label{eq:wick}
	\begin{aligned}
		\big\langle& \tilde\phi_1(k_1) \tilde\phi_1(k_2) \tilde\phi_1(k_3) \tilde\phi_1^*(k_1+k_2+k_3) \big\rangle	
		\\
		&= \big\langle \tilde\phi_1(k_1) \tilde\phi_1(k_2) \big\rangle \,\big\langle \tilde\phi_1(k_3) \tilde\phi_1^*(k_1+k_2+k_3) \big\rangle
		\\
		&\quad + \big\langle \tilde\phi_1(k_1) \tilde\phi_1(k_3) \big\rangle \,\big\langle \tilde\phi_1(k_2) \tilde\phi_1^*(k_1+k_2+k_3) \big\rangle
		\\
		&\quad + \big\langle \tilde\phi_1(k_2) \tilde\phi_1(k_3) \big\rangle \,\big\langle \tilde\phi_1(k_1) \tilde\phi_1^*(k_1+k_2+k_3) \big\rangle .
	\end{aligned}
\end{equation}
Substituting~\eqref{eq:wick} in~\eqref{eq:delta}, and using
\begin{equation}
	\big\langle \tilde\phi_1(k) \tilde\phi_1(k') \big\rangle = \frac{1}{V} \,\frac{\delta_{k,-k'}}{a+3b\bar\phi^2+\kappa k^2} ,
\end{equation}
we then deduce
\begin{equation}
	\begin{aligned}
		\Delta &= \frac{4(T\Delta\mu)^2}{V} \sum_{k_1,k_2} \frac{(k_1k_2)^2 (k_1^2-2k_1k_2)}{(a+3b\bar\phi^2+\kappa k_1^2)(a+3b\bar\phi^2+\kappa k_2^2)}
		\\
		&\quad + {\cal O}(T^3) .
	\end{aligned}
\end{equation}
In the regime $V\gg\Delta x$, for which the sum $\sum_k$ can be approximated by the integral $V\int{\rm d}k/(2\pi)$, we get
\begin{equation}\label{eq:heat_diff}
	\begin{aligned}
		&\frac{\Delta}{4V(T\Delta\mu)^2}
		\\
		&= \iint \frac{k_1^4k_2^2}{(a+3b\bar\phi^2+\kappa k_1^2)(a+3b\bar\phi^2+\kappa k_2^2)} \frac{{\rm d}k_1{\rm d}k_2}{(2\pi)^2} + {\cal O}(T) ,
	\end{aligned}
\end{equation}
where we have simplified the integrand using the $k\to-k$ symmetry.

The range of integration for the wavenumber integral in~\eqref{eq:heat_diff} runs over $[-\pi/\Delta x,\pi/\Delta x]$. In practice, for the discretized version of Active Model B in~\eqref{eq:dyn_disc} taken with the gradient matrix $A_{ij} = (\delta_{i,j-1}-\delta_{i,j+1})/(2\Delta x)$, the odd and even lattice sites decouple when the driving coefficient $\Delta\mu$ is small. In this regime, the field dynamics effectively evolves with a lattice constant $2\Delta x$, so that the appropriate range of integration is then $[-\pi/(2\Delta x),\pi/(2\Delta x)]$: This is the wavenumber integration range that we consider when reporting our analytic prediction~\eqref{eq:heat_diff} in Fig.~\ref{fig2}


\subsection{Motile polar droplets}

Finally, we compute the heat rate for the dynamics of motile polar droplets, as discussed in Sec.~\ref{sec:drop}. The corresponding heat rate is given in~\eqref{eq:entropy_motile}. In the homogeneous state ($\phi_0(x)={\rm cst}$ and $p_0(x)={\rm cst}$), it can be expanded at small noise $T$, yielding at leading order
\begin{equation}\label{eq:entropy_motile_exp_c}
\begin{aligned}
\dot{\cal Q} - V(\dot n_\phi^2 \,&+ \dot n_p^2) 
\\
= T \int_V &\Big[ \big( (\dot n_\phi - \dot n_p) p_0 f_{\phi p} - \dot n_\phi \phi_0 f_{\phi\phi} \big) \big\langle \phi_1 \partial_x p_1 \big\rangle
\\
& + \dot n_\phi \phi_0 \kappa \,\big\langle (\partial_x^2 \phi_1) \,\partial_x p_1 \big\rangle \Big] \,{\rm d}x + {\cal O}(T^2) ,
\end{aligned}
\end{equation}
where we have eliminated some boundary terms.

\begin{figure}
	\centering
	\includegraphics[width=.7\linewidth]{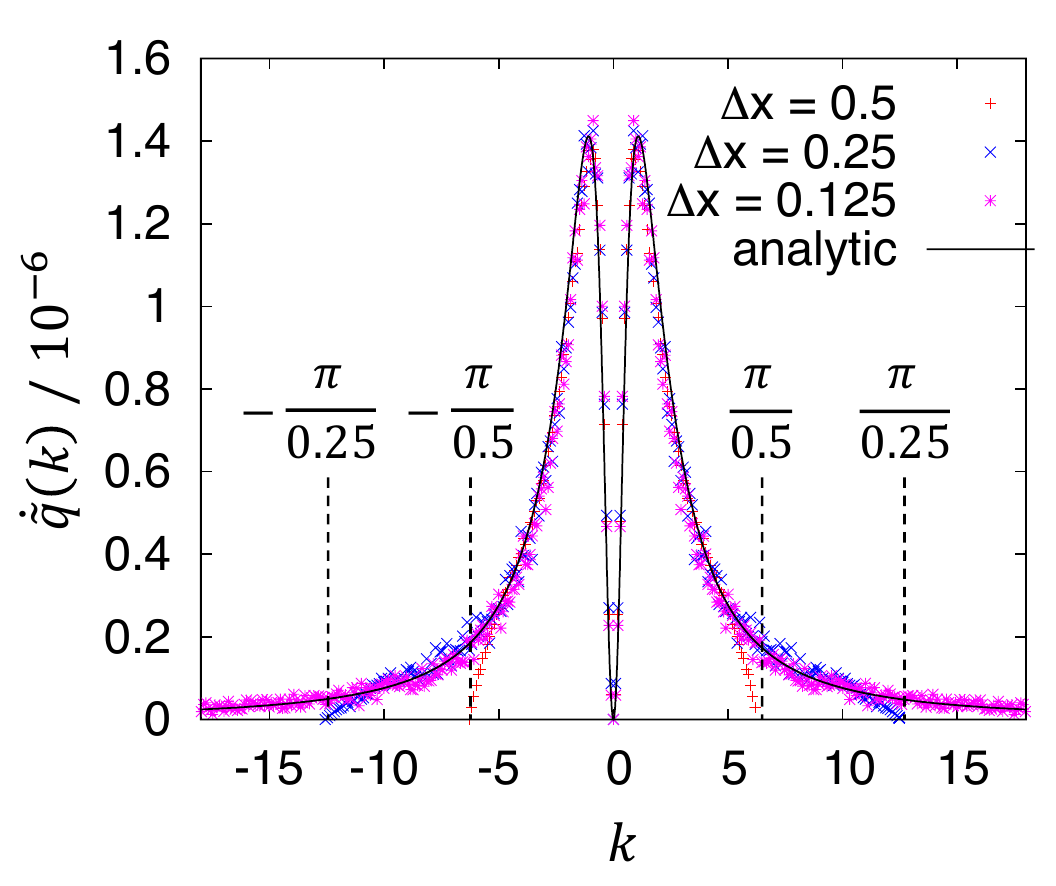}
	\caption{\label{fig6}
		The small-temperature spectrum of heat rate $\dot{\tilde{q}}$, defined by $\dot{\cal Q}-2V\dot n^2 = T \int \dot{\tilde{q}}(k) {\rm d}k + {\cal O}(T^2)$, is measured numerically for different values of lattice spacing $\Delta x$ which controls the upper cut-off $\pi/\Delta x$ of the spectrum. In practice, we express $\dot{\tilde{q}}$ in terms of correlations of the density and polarization fields by taking the Fourier transform of the integrand in~\eqref{eq:entropy_motile_exp_c}. Below the cut-off, we observe a good agreement with our analytical predictions shown in solid lines, as given in~\eqref{eq:spectrum}.
		Simulation details in Appendix~\ref{app:simulations}, same parameter values as in Fig.~\ref{fig4}.
	}
\end{figure}

To evaluate~\eqref{eq:entropy_motile_exp_c}, we compute the correlations between $p_1$ and $\phi_1$ in the Fourier domain. The dynamics of the Fourier coefficients $\big[\tilde\phi_1(k,t), \tilde p_1(k,t)\big] = (1/V) \int_V \big[\phi_1(x,t), p_1(x,t)\big] {\rm e}^{-{\rm i}kx} {\rm d}x$ can be readily deduced from~(\ref{eq:motile_exp}-\ref{eq:motile_exp_b}) as
\begin{equation}\label{eq:dyn_fourier}
\begin{aligned}
\dot{\tilde\phi}_1 &= - \big[ (f_{\phi\phi} + \kappa k^2)k^2 + {\rm i}k\dot n_\phi p_0 \big] \tilde\phi_1
\\
&\quad - (f_{\phi p}k^2 + {\rm i}k\dot n_\phi\phi_0) \tilde p_1 + {\rm i}k \tilde\Lambda_{\phi,0} ,
\\
\dot{\tilde p}_1 &= - (f_{pp} + K k^2 + {\rm i}k\dot n_p p_0) \tilde p_1 - f_{\phi p} \tilde\phi_1 + \tilde\Lambda_{p,0} .
\end{aligned}
\end{equation}
Using It\^o's lemma~\cite{Gardiner}, we obtain the following relations
\begin{equation}\label{eq:ito}
\begin{aligned}
\partial_t \big\langle \tilde p_1\tilde\phi_1^* \big\rangle &= \big\langle \dot{\tilde p}_1\tilde\phi_1^* \big\rangle + \big\langle \tilde p_1 \dot{\tilde\phi}_1^* \big\rangle ,
\\
\partial_t \big\langle |\tilde p_1|^2 \big\rangle &= \big\langle \dot{\tilde p}_1\tilde p_1^* \big\rangle + \big\langle \tilde p_1 \dot{\tilde p}_1^* \big\rangle + 2/V ,
\\
\partial_t \big\langle |\tilde\phi_1|^2 \big\rangle &= \big\langle \dot{\tilde\phi}_1\tilde\phi_1^* \big\rangle + \big\langle \tilde\phi_1 \dot{\tilde\phi}_1^* \big\rangle + 2 k^2/V .
\end{aligned}
\end{equation}
Substituting~\eqref{eq:dyn_fourier} in~\eqref{eq:ito}, we then get in steady state
\begin{widetext}
	\begin{equation}\label{eq:ito_b}
	\begin{aligned}
	0 &= \big[ f_{pp} + (f_{\phi\phi} + K + \kappa k^2)k^2 + i k (\dot n_p - \dot n_\phi)p_0 \big] \big\langle \tilde p_1 \tilde\phi_1^* \big\rangle + f_{\phi p} \big\langle |\tilde\phi_1|^2 \big\rangle + (f_{\phi p}k^2 - {\rm i}k\dot n_\phi\phi_0) \big\langle |\tilde p_1|^2 \big\rangle ,
	\\
	2/V &= 2(f_{pp} + K k^2) \big\langle |\tilde p_1|^2 \big\rangle + f_{\phi p} \big\langle \tilde p_1 \tilde\phi_1^* + \tilde p_1^* \tilde\phi_1 \big\rangle ,
	\\
	2k^2/V &= 2(f_{\phi\phi} + \kappa k^2) k^2 \big\langle |\tilde\phi_1|^2 \big\rangle + f_{\phi p} k^2 \big\langle \tilde p_1 \tilde\phi_1^* + \tilde p_1^* \tilde\phi_1 \big\rangle + {\rm i}k\dot n_\phi \phi_0 \big\langle \tilde p_1 \tilde\phi_1^* - \tilde p_1^* \tilde\phi_1 \big\rangle .
	\end{aligned}
	\end{equation}
	From~\eqref{eq:ito_b}, we obtain the solution for $\langle \tilde p_1 \tilde\phi_1^* \rangle$, and, after substituting in~\eqref{eq:entropy_motile_exp_c}, we get the explicit expression of $\dot{\cal Q}$. For $\dot n_\phi=\dot n_p\equiv\dot n$, it takes the following form
	\begin{equation}\label{eq:spectrum}
	\dot{\cal Q} = 2V\dot n^2 + T V \int_{-\pi/\Delta x}^{\pi/\Delta x} \frac{(\dot n_\phi \phi_0 k)^2 (f_{\phi\phi} + \kappa k^2)^2 \big[ f_{pp} + (f_{\phi\phi} + K + \kappa k^2)k^2 \big] }{ \big[ (f_{\phi\phi} + \kappa k^2)(f_{pp} + K k^2) - f_{\phi p}^2 \big] \big[ f_{pp} + (f_{\phi\phi} + K + \kappa k^2)k^2 \big]^2 - (\dot n_\phi\phi_0 f_{\phi p})^2} \frac{{\rm d}k}{2\pi} + {\cal O}(T^2) .
	\end{equation}
\end{widetext}
We have measured numerically the spectrum of the integrand in the right-hand side of~\eqref{eq:entropy_motile_exp_c}, where $\sqrt T\{\phi_1,p_1\}$ are replaced by $\{\phi-\phi_0,p-p_0\}$, respectively: Fig.~\ref{fig6} shows a good agreement with our prediction in~\eqref{eq:spectrum}.


\section{Numerical simulations}\label{app:simulations}

All numerical simulations are performed at dimension $d=1$ in a box of size $V$ with periodic boundary conditions at $x=0$ and $x=V$. We discretize time and space into $t=m\Delta t$ and $x=i\Delta x$, where $m=\{0,1,2,\dots\}$, $i=\{0,1,\dots,N-1\}$, and $V=N\Delta x$.

The numerical scheme for Sec.~\ref{sec:phase} is the same as in~\cite{Nardini2017}. In particular, it corresponds to choosing the discretization $C_i^{(1)}$ with $A_{ij} = (\delta_{i,j-1}-\delta_{i,j+1})/(2\Delta x)$ in~\eqref{eq:Cphi-discretized}, so that the spurious drift terms vanish. The time and spatial discretization constants are fixed to be $\Delta t=0.01$ and $\Delta x=1$. The numerical scheme for Sec.~\ref{sec:drop} is as follows. We assume the fields $\{\phi^m_i,p^m_i\}$ live on-lattice with periodic boundary condition:
\begin{equation}
	\phi_{0}^{m} = \phi_{N-1}^{m} ,
	\quad
	p_{0}^{m} = p_{N-1}^{m},
\end{equation}
whereas the current $J_{i+\frac{1}{2}}^{m}$ and the conserved-noise $\Lambda_{\phi,i+\frac{1}{2}}^{m}$ live off-lattice. This ensures the fields $\{\phi^m_i,p^m_i\}$ at the odd and even sites are coupled even in the small activity regime. At each time step $m$, the current is evaluated as
\begin{equation}
	\begin{aligned}
		J_{i+\frac{1}{2}}^{m} & =-\frac{1}{\Delta x}\left[\left(\frac{\delta\mathcal{F}}{\delta\phi}\right)_{i+1}^{m}-\left(\frac{\delta\mathcal{F}}{\delta\phi}\right)_{i}^{m}\right]
		\\	
		&\quad +\dot{n}_\phi\frac{\phi_{i+1}^{m}p_{i+1}^{m}+\phi_{i}^{m}p_{i}^{m}}{2}+\sqrt{\frac{2T}{\Delta t\Delta x}} \Lambda_{\phi,i+\frac{1}{2}}^{m},\label{eq:J-discretized}
	\end{aligned}
\end{equation}
and the fields are updated according to
\begin{equation}\label{eq:phidot-discretized}
	\phi_{i}^{m+1} = \phi_{i}^{m} - \frac{\Delta t}{\Delta x} \Big[ J_{i+\frac{1}{2}}^{m}-J_{i-\frac{1}{2}}^{m} \Big] ,
\end{equation}
and
\begin{equation}\label{eq:pdot-discretized}
	\begin{aligned}
		p_{i}^{m+1} &= p_{i}^{m}+\Delta t \Bigg[ \left(\frac{\delta\mathcal{F}}{\delta p}\right)_{i}^{m} - \dot{n}_p\frac{(p_{i+1}^{m})^2 - (p_{i-1}^{m})^2}{4\Delta x}  \Bigg]
		\\
		&\quad +\sqrt{\frac{2T\Delta t}{\Delta x}}\Lambda_{p,i}^{m},
	\end{aligned}
\end{equation}
where $\{\Lambda_{\phi,i+\frac{1}{2}}^{m},\Lambda_{p,i}^{m}\}$ are Gaussian random variables with zero mean and unit variance, independent for each $i$ and $m$. Note that we have chosen discretization $C_{p,i}^{(2)}$ from~\eqref{eq:Cp-discretized} so that the spurious drift terms vanish.

The multiplying factor in the last term of~\eqref{eq:J-discretized} and~\eqref{eq:pdot-discretized} comes from the regularization of the delta function in~\eqref{eq:lambda}. The conserved noise $\Lambda_{\phi,i+\frac{1}{2}}^m$ lives off-lattice, which means that its values are specified only for half-integer lattice sites, whereas the non-conserved noise $\Lambda_{p,i}^m$ lives on-lattice. The numerical results above does not depend strongly on $\Delta t$ (we choose $\Delta t=10^{-3}$ to $10^{-5}$), however it depends slightly on the spatial discretization $\Delta x$, as shown in Fig.~\ref{fig6}.

The average density and polarization profiles in Figs.~\ref{fig5}(a,b) and the heat rate profiles in Figs.~\ref{fig5}(c,d) are computed in the co-moving frame of the droplets, since the droplets are moving with constant velocity towards $x>0$. First we compute the centre of mass of the droplet(s) $R(t)=\int x\phi(x,t)\,{\rm d}x\big/\int \phi(x,t)\,{\rm d}x$. The instantaneous density profile in the co-moving frame will then be $\phi(x,t)\rightarrow\phi(x-R(t),t)$. Fig.~\ref{fig5}(a) is the long-time average of $\phi$ in the co-moving frame: $\left<\phi(x)\right>=\lim_{t_1\rightarrow\infty}\int_0^{t_1}\phi(x-R(t),t)\,{\rm d}t/t_1$. Similar procedure is performed for the average polarization $\left<p(x)\right>$ (Fig.~\ref{fig5}(b)) and heat rate profile $\dot{q}(x)$ (Figs.~\ref{fig5}(c,d)).


%

\end{document}